\documentclass[11pt,a4paper]{article}
\usepackage{jcappub}

\usepackage{hyperref}
\usepackage{graphicx}
\usepackage{subcaption}
\graphicspath{{./picture/}}
\usepackage{empheq}
\usepackage{amssymb}
\usepackage{mathtools}
\usepackage{commath}
\usepackage{verbatim}
\usepackage{xcolor}
\usepackage{soul}
\usepackage{booktabs}
\usepackage{amsmath}
\usepackage{hhline}
\usepackage[scale={0.75,0.8}]{geometry}
\usepackage{url}
\usepackage{caption}
\usepackage[numbers, sort&compress]{natbib}
\usepackage{epsfig}
\usepackage{lscape}
\usepackage{mathrsfs}
\usepackage{nicefrac} 
\usepackage{upgreek}
\usepackage{bbm}
\usepackage{psfrag}
\usepackage{hyperref}

\newcommand{\varphio}{\bar{\varphi}}

\DeclareMathAlphabet{\mathpzc}{OT1}{pzc}{m}{it}

\newcommand{\ii}{i}

\newcommand{\corr}[1]{\textcolor{black}{#1}}
\newcommand{\corrmaybe}[1]{\textcolor{black}{#1}}

\newcommand{\n}{{\rm n}}

\newcommand{\kk}{{\rm k}}

\newcommand{\Deltak}{\Delta{\rm k}}

\usepackage{hyperref}
\newcommand{\g}{\mathsf{g}}
\newcommand{\wilson}{{\rm c}}

\newcommand{\SM}{\mathsf{\corr{a}}}
\newcommand{\epsilonvar}{\varepsilon_\varphi}
\newcommand{\epsilonvarM}{\varepsilon_{pl}}

\newcommand{\mass}{\mathcal{M}}
\newcommand{\M}{M}
\newcommand{\f}{f}
\newcommand{\Minf}{{\rm M}}
\newcommand{\fluct}{\phi}
\newcommand{\GG}{\Gamma_\varphi}

\newcommand{\disprel}{\mathcal{W}}
\newcommand{\W}{\Upomega}
\newcommand{\w}{\Omega}
\newcommand{\V}{\mathcal{V}}
\newcommand{\Vend}{\mathcal{V}_{\rm end}}
\newcommand{\Treh}{T_{\rm re}}
\newcommand{\x}{\textbf{x}}

\newcommand{\rhoend}{\rho_{\rm end}}
\newcommand{\rhoreh}{\rho_{\rm re}}

\newcommand{\wrehbar}{\bar{w}_{\rm re}}
\newcommand{\wreh}{w_{\rm re}}
\newcommand{\Nreh}{N_{\rm re}}
\newcommand{\X}{\mathcal{X}}
\newcommand{\vv}{\mathsf{v}}

\title{Measuring the Inflaton Coupling in the CMB} 

\affiliation{Centre for Cosmology, Particle Physics and Phenomenology, Universit\'{e} catholique de Louvain, Louvain-la-Neuve B-1348, Belgium}
\author{Marco Drewes}
\abstract{We study the perspectives to extract information about the microphysical parameters that governed the reheating process after cosmic inflation from CMB data. We identify conditions under which the inflaton coupling to other fields can be constrained for a given model of inflation without having to specify the details of the particle physics theory within which this model is realised. This is possible when the effective potential during reheating is approximately parabolic, and when the coupling constants are smaller than an upper bound that is determined by the ratios between the inflaton mass and the Planck mass or the scale of inflation. We consider scalar, Yukawa, and axion-like interactions and estimate that these conditions can be fulfilled if the inflaton coupling is comparable to the electron Yukawa coupling or smaller, and if the inflaton mass is larger than $10^5$ GeV. Constraining the order of magnitude of the coupling constant requires measuring the scalar-to-tensor ratio at the level of $10^{-3}$, which is possible with future CMB observatories. Such a measurement would provide an important clue to understand how a given model of inflation may be embedded into a more fundamental theory of nature.}
\newpage

\begin{document}
\maketitle

\section{Introduction}

\emph{Cosmic inflation} \cite{Starobinsky:1980te,Guth:1980zm,Linde:1981mu} is currently the most popular explanation for both, the overall homogeneity and isotropy of the observable universe, and the 
observed correlations amongst the small cosmological perturbations that are visible in the Cosmic Microwave Background (CMB) and formed the seeds for galaxy formation. 
While the general idea that he universe underwent a phase of accelerated expansion before it entered the radiation dominated epoch provides an excellent explanation for the observed data, it is unclear what mechanism drove this expansion,\footnote{A partial overview of scenarios can e.g. be found in ref.~\cite{Martin:2013tda}.} and how it should be embedded into a more fundamental theory of nature.
Given the high energy scales that most inflationary models invoke, it seems unlikely that the mechanism of inflation and its connection to particle physics theories can be directly probed in laboratory experiments in the foreseeable future.\footnote{A noteworthy exception from this rule is described in ref.~\cite{Bezrukov:2009yw}.} 
An indirect probe of this connection can, however, be obtained by studying the impact that the production of particles after inflation, known as \emph{cosmic reheating}~\cite{Albrecht:1982mp,Abbott:1982hn,Dolgov:1989us,Traschen:1990sw,Shtanov:1994ce,Kofman:1994rk,Boyanovsky:1996sq,Kofman:1997yn}, has on cosmological observables.
It is well-known that this process leaves an imprint in CMB observables through its effect on the expansion history
\cite{Martin:2010kz,Adshead:2010mc,Easther:2011yq},\footnote{A more direct messenger could potentially be provided by gravitational waves \cite{Caprini:2018mtu}, cf.~also \cite{Ghiglieri:2020mhm,Ringwald:2022xif}.} 
which is primarily governed by the  equation of state parameter $\wreh$ during reheating, and by the duration of the reheating process in terms of $e$-folds $\Nreh$.
In \cite{Drewes:2015coa} it was pointed out that 
this effect can be used to constrain ``fundamental" (microphysical) parameters, in particular the inflaton coupling(s) $\g$ to other fields. This connection has been explored quantitatively for $\alpha$-attractors in 
\cite{Ueno:2016dim,Drewes:2017fmn,Ellis:2021kad}.
However, translating a constraint on $\Nreh$ into a constraint on $\g$ relies on the assumption that a relation between $\g$ and CMB observables can be established without fixing other details of the underlying particle physics model. 
In the present work we outline general conditions under which this is justified.

We focus on the simplest scenarios, in which the energy density during inflation is dominated by the potential energy density $\V(\varphi)$ of the condensate $\varphi=\langle\Phi\rangle$ of a single scalar field $\Phi$, but it is straightforward to generalise the considerations to multi-field inflation. 
 To phrase the goal of this work more precisely, we distinguish between the following sets of microphysical parameters.
\begin{itemize}
\item \textbf{Model of inflation}: \corr{A} model on inflation is defined by the choice of the effective potential $\V(\varphi)$. $\V(\varphi)$ is specified by the complete set of the inflaton's self-interactions (i.e., operators that are constructed from $\Phi$ only) with coupling constants $\{\vv_i\}$. 
\item \textbf{Inflaton couplings}:
\corr{W}e refer to all terms in the action that contain both, $\Phi$ and other fields, as \emph{inflaton couplings}. 
Their strength is determined by a set of coupling constants $\{\g_i\}$.
\item \textbf{Full particle physics model}: \corr{A} given model of particle physics is typically defined by its field content and its symmetries. 
In general it contains a much larger set of parameters than the combined sets $\{\vv_i\}$ and $\{\g_i\}$, including  the masses of the particles produced during reheating as well as their interactions amongst each other and with all other particles. We refer to the set of all parameters other than $\{\vv_i\}$ and $\{\g_i\}$ as $\{\SM_i\}$. This set e.g.~includes the parameters of the SM.
\end{itemize}
We shall in the following assume that a model of inflation has been chosen, i.e., $\V(\varphi)$ has been specified in terms of the $\{\vv_i\}$. We further assume that the numerical values of the $\{\vv_i\}$ are either fixed by theoretical considerations or can be extracted from data.
Since the number of cosmologically observable parameters is quite limited (here we consider the spectral index $n_s$ of scalar CMB perturbations, their amplitude $A_s$, and the scalar to tensor ratio $r$), meaningful constraints on any individual microphysical parameter can only be derived if $\Nreh$  depends on just a small number of these parameters. 
In the present work we address the following question. 
Suppose the inflaton reheats the universe through a particular type of interaction that is governed by a dimensionless coupling constant $\g \in \{\g_i\}$, what is the range of values for $\{\vv_i\}$ and  $\g$ 
 for which these parameters can be constrained from CMB data independently of the underlying particle physics model?

\section{
Connecting inflation to particle physics
}
\label{Section:ConnectionInflationToParticle}
For the present discussion we assume that the time evolution of $\varphi$ can be described by an equation of the form\footnote{
We comment on the justification to use an equation of the form \eqref{EOM} in Appendix \ref{ApendixQFT}.
}
\begin{equation}\label{EOM}
    \ddot{\varphi} + (3H + \GG)\dot{\varphi} + \partial_\varphi \V(\varphi)=0
\end{equation}
\subsection{The impact of reheating on the CMB}
The the primordial spectrum of perturbations before reheating is determined by $\V(\varphi)$ 
and therefore depends only on the $\{\vv_i\}$.
If the transition from inflation to the radiation dominated epoch were instantaneous, then 
this would entirely fix the CMB power spectrum 
(leaving aside foreground effects,
and assuming a standard thermal history after inflation).
However, 
reheating takes a number $\Nreh$ of $e$-folds
and should be regarded as a separate \emph{reheating epoch} in cosmic history. This epoch begins when the universe stops accelerating (the equation of state exceeds $w>-1/3$) and ends when the energy density of radiation\footnote{
 We collectively refer to all relativistic particles as ``radiation", irrespectively of their spin and charges. 
 }
$\rho_R$ exceeds the energy density $\rho_\varphi\simeq \dot{\varphi}^2/2 + \V(\varphi)$ stored in $\varphi$, i.e., the universe becomes radiation dominated ($w=1/3$).
The modified equation of state parameter $\wreh$ during reheating affects the redshifting of cosmological perturbations and thereby leaves an imprint in the CMB \cite{Martin:2010kz,Adshead:2010mc,Easther:2011yq}.
Since we can only observe the time-integrated effect, the quantities that the CMB is directly sensitive to are the duration of the reheating era in terms of $e$-folds $N_{\rm re}$ and the averaged equation of state $\wrehbar$ during reheating, with
\begin{equation}
\wrehbar= \frac{1}{N_{\rm re}}\int_0^{N_{\rm re}} w(N) dN.
\end{equation}
Within a given model of inflation, all parameters that affect the CMB power spectrum except $\Nreh$  are fixed by the $\{\vv_i\}$ alone, as $\V(\varphi)$ permits to compute the initial spectrum of cosmological perturbations,
fixes the moment when inflation ends ($\partial_\varphi\V(\varphi)>3H\dot{\varphi}$) and in good approximation determines $\wreh$ (since the energy density is by definition dominated by $\varphi$ during reheating). 
The impact of $\Nreh$ can be expressed in terms of the energy density at the end of reheating $\rho_{\rm re}$.
 In practice $\rhoreh$ is often parameterised in terms of an effective temperature of the plasma formed by the decay products, the \emph{reheating temperature}
\begin{equation}\label{rhoreeq}
    \frac{\pi^2 g_*}{30}T_{\rm re}^4\equiv \rhoreh = \rhoend\exp(- 3\Nreh(1 + \wrehbar)),
\end{equation}
with $\rhoend$ the energy density at the end of inflation.
At first sight the dependence on $\Nreh$ is quite inconvenient, as it poses a fundamental uncertainty in the 
fitting of the $\{\vv_i\}$ from CMB data. 
It is, however, possible to turn the tables and use this dependence to constrain $\Nreh$ and $T_{\rm re}$ from CMB data for a given model $\V(\varphi)$ by fitting $\Nreh$ to CMB data. 
Such kind of constraints have been derived for a large number of models, including
Starobinski inflation \cite{Cook:2015vqa},
$\alpha$-attractor inflation \cite{Eshaghi:2016kne,Ueno:2016dim,Nozari:2017rta,DiMarco:2017zek,Drewes:2017fmn,Maity:2018dgy,Rashidi:2018ois,Mishra:2021wkm,Ellis:2021kad}, 
natural inflation \cite{Munoz:2014eqa,Cook:2015vqa,Wu:2018vuj,Stein:2021uge},
power law \cite{Cai:2015soa,DiMarco:2018bnw,Maity:2018qhi,Maity:2019ltu,Antusch:2020iyq} and polynomial potentials \cite{Dai:2014jja,Cook:2015vqa,Domcke:2015iaa,Dalianis:2016wpu},
Higgs inflation \cite{Cook:2015vqa,Cai:2015soa},
curvaton models \cite{Hardwick:2016whe},
hilltop type inflation \cite{Cook:2015vqa,Cai:2015soa},
axion inflation \cite{Cai:2015soa,Takahashi:2019qmh},
inflection point inflation \cite{Choi:2016eif},
fiber inflation \cite{Cabella:2017zsa},
tachyon inflation \cite{Nautiyal:2018lyq}, 
K\"ahler moduli inflation  \cite{Kabir:2016kdh,Bhattacharya:2017ysa},
and other SUSY models \cite{Cai:2015soa,Dalianis:2018afb}.
In most of these $\Nreh$ (or equivalently $T_{\rm re}$) is practically treated as a free fit parameter, without relating it to microphysics. In the present article we investigate the general conditions under which information about $\Nreh$ obtained from the CMB can be translated into constraints on individual microphysical parameters (specifically $\g$).

\subsection{The dependence of CMB observables on microphysical parameters}\label{ParameterDependencies}
Reheating ends and $\rho_\varphi$ is rapidly converted into radiation once $\GG$ exceeds the Hubble \corr{rate} $H$.
As far as the effect on the expansion history is concerned this occurs almost instantaneously \cite{Martin:2010kz}.  
We work in this approximation in the following, which permits to establish the well-known relation between $\Treh$ or $\rhoreh$ and $\GG$,
\begin{equation}
\frac{\pi^2 g_*}{30}T_{\rm re}^4= \rho_R\big|_{\GG=H} = \rho_{\rm re} \,.
\quad {\rm as} \quad
T_{\rm re}=\sqrt{\GG M_{pl} }\left(\frac{90}{\pi^2g_*}\right)^{1/4}\Big|_{\GG=H},\label{TR}
\end{equation}
with $g_*$  the effective number of relativistic degrees of freedom and $M_{pl}=2.435\times 10^{18}~{\rm GeV}$  the reduced Planck mass.
 Via the Friedmann equation  
  \begin{equation}\label{Friedmann}
     H^2=\frac{\rho}{3  M_{pl}^2} \quad {\rm with} \quad
     \rhoend\simeq\frac{4}{3} \Vend \equiv \frac{4}{3} \V(\varphi_{\rm end})
 \end{equation}
 (with $\varphi_{\rm end}$ the value of $\varphi$ at the end of inflation),
  the condition $\GG=H$ can be used to translate a bound on $\Nreh$ into a constraint on the damping rate $\GG$, 
 \begin{eqnarray}
\GG|_{\GG=H}=\frac{1}{ M_{pl}}\left(\frac{\rhoend}{3}\right)^{1/2} e^{-3(1+\wrehbar) \Nreh/2}\, \label{GammaConstraint}
\end{eqnarray}
Within a given model of inflation, all parameters on the right hand side of eq.~\eqref{GammaConstraint} can be constrained with CMB data.
The details of this connection are not crucial for the present work, 
a sketch of the derivation  
and an explicit expression for 
$\Nreh$ in terms of observable\corr{s} 
are given appendix \ref{RelationToObservables}. 
Our task here is to evaluate the conditions under which \eqref{GammaConstraint} can be translated into a measurement of $\g$. 
In the simplest case $\GG$ depends only on the $\{\vv_i\}$ and $\g$ (e.g.~if reheating is driven by elementary decays of inflaton quanta into light particles in vacuum, cf.~table~\ref{ThermalRates}), 
meaning that a constraint on $\g$ can directly be read off \eqref{GammaConstraint}. 
However, in practice the relation between $\g$ and 
cosmological observables is usually complicated and depends on the underlying particle physics model (and therefore the $\{\SM_i\}$) in several ways.
\begin{itemize}
    \item \textbf{Feedback effects on $\GG$.} 
    At the end of inflation the universe is practically empty, but as soon as reheating commences, it becomes filled with particles. 
    These particles affect $\GG$ even if the energy density is still dominated by $\varphi$, e.g.~by triggering induced transitions or by scattering with the quanta that form the condensate $\varphi$.
    The feedback not only poses a practical challenge by making the reheating process highly non-linear and difficult to simulate \cite{Amin:2014eta}, 
    it also makes it impossible to constrain $\g$ independently of the particle physics model, as feedback necessarily depends on the properties of the produced particles and their occupation numbers.

\item \textbf{Number of degrees of freedom.}
The relation between $\Nreh$ and the CMB parameters $(A_s, n_s, r)$ depends logarithmically on
the effective numbers of degrees of freedom
$g_*$ and $g_{s*}$ 
related to the plasma's energy and entropy densities at the time of reheating, \corr{respectively}, cf.~\eqref{Nre}. 
These parameters can only be fixed when the complete field content of the particle physics model is specified.
At first sight this makes it impossible to constrain $\g$ without specifying this model. 
However, the resulting dependency is quite weak.
Plugging \eqref{Nre} 
into \eqref{GammaConstraint} yields
$\GG \propto (g_{s*}^{1/3}/g_*^{1/4})^{6(1+\wrehbar)/(1-3\wrehbar)}$,
For $g_{s*}=g_*$ and $\wrehbar = 0$ this reduces to $\GG \propto \sqrt{g_*}$. Assuming $\GG\propto \g^2$ this only leads to a very mild dependence $\g\propto g_*^{1/4}$, which leads to an uncertainty that is well below the sensitivity of near future CMB missions unless $g_*$ changed by several orders of magnitude.

   \item \textbf{Plasma equilibration.}
   $T_{\rm re}$ can be interpreted as a physical temperature only under the assumption that the bath 
 reaches thermal equilibrium instantaneously. 
   The equilibration affects the present discussion insofar as that  deviations of the produced particles' phase space distributions from equilibrium \corr{can} lead to deviations from $w=1/3$, and that it affects the relation between the right-hand side of \eqref{GammaConstraint} and the CMB parameters, cf.~equation (15) and following in \cite{Ueno:2016dim}. 
 A recent discussion of the equilibriation can e.g.~be found in~\cite{Mazumdar:2013gya,Harigaya:2013vwa,Harigaya:2014waa,Mukaida:2015ria}. In typical UV-completions  of the SM that contain coupling constants of order one the resulting error is small, hence the assumption of fast thermalisation only imposes a comparably weak condition on the $\{\SM_i\}$.
 
  \item \textbf{Standard expansion history.}
Throughout this work we assume a standard expansion  history, i.e., radiation domination ($w=1/3$) between the end of reheating and \corr{big bang nucleosynthesis} (BBN).\footnote{ 
A critical discussion of this assumption can be found in \cite{Allahverdi:2020bys}.}
If e.g.~entropy was released into the plasma after reheating, then back-extrapolation of the present CMB temperature leads to an overestimate of the temperature before the moment of the release.
Hence, if an unknown amount of entropy is injected after reheating, one can only obtain an upper bound on $\g$ from the CMB.
  \item \textbf{Gravitational waves.}
Tensor perturbations can also be caused by gravitational waves generated after inflation, and this can affect the observed value of $r$.\footnote{Observationally they may e.g.~be spotted due to deviations from the relation $r+8n_t=0$ between $r$ and the spectral index of tensor perturbations $n_t$, as observations are in principle able to distinguish different sources of tensor modes \cite{Hiramatsu:2018nfa}.}
If the gravitational waves are caused by the inflationary sector (e.g.~they were generated during reheating), their effect on $r$ can at least in principle be computed with knowledge of the $\{\vv_i\}$, otherwise this brings in dependencies on the $\{\SM_i\}$.
  In models where $r$ is dominated by contributions that were generated after inflation and that are not calculable from knowledge of the $\{\vv_i\}$,  one cannot use $r$ for the approach proposed here. One may, however, still utilise $n_s$ (and potentially future measurements of its running).
  \item \textbf{Foreground and late time effects.}
  Assuming that all astrophysical foregrounds have been identified and removed from the CMB data, 
  there can be cosmological effects that are not part of the concordance model. 
These can e.g.~be caused by comparably light new particles at later times, which can modify CMB observables in different ways.  
On one hand, they can affect the propagation of photons after their decoupling, e.g.~through cosmic birefringence caused by a modification of Maxwell's equation (cf.~\cite{Komatsu:2022nvu} and reference therein), 
On the other hand, they can modify the interactions of particles before photon decoupling \cite{Tram:2016rcw}. A well-studied example are neutrino non-standard interactions \cite{Oldengott:2017fhy,Barenboim:2019tux,RoyChoudhury:2022rva}.
The existence of light new particles is strongly constrained by cosmology and experiments. 
If they are light during photon decoupling they  introduce deviations from the SM prediction for the effective number of relativistic species $N_{\rm eff}^{\rm SM}=3.0440\pm0.0002$ \cite{Bennett:2020zkv}  (cf.~also \cite{Akita:2020szl,Froustey:2020mcq,EscuderoAbenza:2020cmq})
 which are strongly constrained by observations \cite{Planck:2018vyg,BICEP:2021xfz}.
 If they are heavier, there are constraints from experiment and astrophysics, cf.~\cite{Agrawal:2021dbo} for a review.
The approach proposed here is not applicable to specific models that can avoid all of these constraints.
\item \textbf{Instantaneous decay.}
We  assume that the conversion of the remaining inflaton energy density $\rho_\varphi$ into $\rho_R$ occurs instantaneously when $\GG=H$. While this is usually a good approximation, relaxing it introduces a mild dependence on $g_*$ during the transitional phase, which we ignore here.
 \item \textbf{
 Radiative corrections.}
We define the $\{\vv_i\}$ as the microphysical parameters that characterise the effective potential during reheating. 
The CMB anisotropies are generated during the slow roll phase (which is sensitive to a different part of $\V(\varphi)$) and do not directly probe the shape of the potential near its minimum. 
Since inflation and reheating can occur at different energy scales, a fully consistent determination of the $\{\vv_i\}$ may require solving renormalisation group equations that in principle bring in a dependence on the $\{\SM_i\}$. 
Radiative corrections can further affect the relation between \corr{the} parameters \corr{in} the Lagrangian and the properties of \corr{the} perturbations generated during inflation in models where the field space trajectories during inflation can be sensitive to small corrections (e.g.~inflection point inflation).
In section \ref{sec:rangeofmeasurable} we find that a determination of $\g$ from the CMB is only possible for comparably small values of the inflaton coupling constants, which suggests that these effects are subdominant in the regime where the methods discussed here can be applied \cite{Herranen:2013raa,Herranen:2016xsy}. 
\item \textbf{Multifield effects.}
Many single field models of inflation are motivated by
particle physics theories that predict the existence of multiple scalar fields.
During inflation one can justify the single field description by identifying the inflaton with the relevant direction in field space. 
It is in general not clear whether this can be justified during reheating, cf.~e.g.~\cite{Martin:2021frd}.
There are at least two qualitatively different effects that can arise from this even if we assume that all other scalar fields have relaxed to their minimal before or during inflation.
Firstly, quantum fluctuations \cite{Linde:1982uu,Starobinsky:1986fx} of light spectator fields can build up on  super-horizon scales (cf.~\cite{Moss:2016uix} and references therein for a recent discussion) and locally displace those fields away from the minima of their potential. 
They may either get trapped in false vacua that subsequently decay or oscillate around the true minima. 
This can affect the thermal history during reheating \cite{Passaglia:2021upk}. These effects strongly depend on the field content of the theory in which a given model. Based on the considerations in appendix \ref{ThermalFeedbackAppendix} it seems that they are unlikely to affect $\Nreh$ unless the energy stored in these degrees of freedom dominates the universe. 
Secondly, in models that feature kinetic mixing between the fields, a non-trivial metric in field space that can introduce instabilities \cite{Renaux-Petel:2015mga}  that lead to violent particle production.
This is, for instance, well-known to occur in $\alpha$-attractor models \cite{Krajewski:2018moi,Iarygina:2018kee,Iarygina:2020dwe} and in models with non-minimal coupling to gravity \cite{DeCross:2015uza,Ema:2016dny} (including Higgs inflation \cite{Sfakianakis:2018lzf}), where the non-trivial field space metric is introduced by the mapping between the Einstein and Jordan frame.\footnote{ If the mixing occurs only with other fields in the (multi-field) inflaton sector, then $\GG$ may formally still be independent of the $\{\SM_i\}$ (because the parameters in the extended inflaton sector belong to the set $\{\vv_i\}$), though the determination of its functional form  would be practically difficult due to highly non-linear and the non-perturbative behaviour. In practice the explosive particle production will almost always introduce $\{\SM_i\}$-dependencies through subsequent particle decays and scatterings.}
In the present work we focus on phenomena that can be described within the single field framework, additional conditions that arise from specific embeddings in multifield models may exist, but have to be studied in each model separately.
\end{itemize}
To summarise, within the set of models where our approach is in principle applicable, 
the main complication arises from feedback effects of the produced particles on the dissipation rate $\GG$. All other effects are either subdominant in the regimes identified in section \ref{sec:rangeofmeasurable}, or only lead to mild restrictions on the properties of the underlying particle physics theories. 
We shall therefore focus on the conditions that can be derived from the requirement to avoid feedback.

\section{Conditions to measure the inflaton coupling}\label{SecConditions}
Feedback typically becomes relevant when the occupation numbers in the plasma approach unity. One can roughly distinguish two effects.
\begin{itemize}
    \item[1)] \emph{Quantum statistical effects.} When the occupation numbers of individual modes in which particles are being produced reach unity, quantum statistical effects considerably modify the particle production rate for these modes. Depending on the spin of the final state particles, this either enhances $\GG$ by induced transitions or suppresses the rate due to Pauli blocking.
    \item[2)] \emph{Plasma effects.} When the overall density of particles in the plasma becomes too large, then scatterings between the (quasi)particles that make up the plasma and those that form $\varphi$ become frequent and modify $\GG$.
\end{itemize}
In the following we restrict the discussion to scenarios where the cosmos is reheated during oscillations of $\varphi$ around its potential minimum. 
Other forms of reheating, such as tachyonic instabilities, are generally driven by non-perturbative processes that introduce a complicated dependence of $\GG$ on time and the $\{\SM_i\}$.
In this situation, one can broadly distinguish the following mechanisms to reheat the universe,
\begin{itemize}
\item[(i)] elementary decays of inflaton quanta, 
\item[(ii)] annihilations of inflaton quanta,
\item[(iii)] scatterings of inflaton quanta with other particles.
\item[(iv)] non-perturbative production of particles from the time-dependent background provided by $\varphi$,
\item[(v)] gravitational effects.
\end{itemize}
We shall neglect (v) in the following, as it tends to be sub-dominant when other channels exist.\footnote{
For $\g$ within the range permitted by the considerations in section \ref{sec:rangeofmeasurable},
a pure coupling to gravity typically leads to fragmentation \cite{Easther:2010mr} before the end of the reheating epoch (cf.~\cite{Drewes:2014pfa,Ming:2021eut} for analytic estimates). However, as long as this mainly transforms the coherent condensate $\varphi$ into a plasma of non-relativistic $\Phi$-particles that decay perturbatively this should not change our conclusions. The reason is that the perturbative decay rate of the latter equals the decay rate $\GG$ of the condensate at leading order when non-linear effects can be neglected.} 
The mechanisms (i), (ii) and (iv) all can potentially generate large occupation numbers in specific momentum modes well before the overall density of the produced particles becomes large. We shall therefore discuss feedback of type 1) first. 
It is convenient to introduce the small parameters
\begin{equation}\label{Smallness}
\epsilonvarM \equiv m_\varphi/M_{pl} \ ,\quad \epsilonvar \equiv m_\varphi/\varphi_{\rm end}.
\end{equation}

\subsection{Restrictions on the potential: qualitative discussion}
We will see in section \ref{sec:selfinteractions} that the requirement to avoid a so-called broad resonance due to the non-perturbative particle production (iv) practically 
restricts us to scenarios where $\varphi_{\rm end}$ is small enough that $\V(\varphi)$ can be approximated by a parabola, i.e., the mildly non-linear regime where reheating can be modelled by a  harmonic oscillator with time dependent frequency and damping. 
We can intuitively understand this point already without diving too deeply into computations by 
using the adiabaticity condition
on the dispersion relations $\disprel_\phi$ for screened $\phi$-particles
\begin{equation}\label{adiabaticity}
    \frac{\partial_t \corr{\disprel}_\phi}{\corr{\disprel}_\phi^2} \ll 1,
\end{equation}
which we use as a rough estimator for the presence of highly efficient non-perturbative particle production in a so-called \emph{broad resonance}.
The full dispersion relation $\disprel_\phi$ is defined as a pole of the resummed propagator,\footnote{To be precised, we define $\disprel_\phi$ as the real part of the pole of the spectral function, i.e., the Fourier transform of the spectral propagator $\Delta_\phi^-$ defined in \eqref{SpectralStatistical} with respect to $x_1 - x_2$.} 
in a simple scalar field theory it can be approximated as $\disprel_\phi\simeq \W_\phi^2 =\mathbf{k}^2 + \mass_\phi^2$, with $\mass_\phi$ an effective mass that in general depends on $\varphi$ and the properties of the screening plasma. 
For vanishing field value it reduces to 
$\w_\phi^2 = {\bf k}^2 + \M_\phi^2  = \W_\phi^2|_{\varphi=0}$, with $\M_\phi$ a thermal mass that \corr{includes corrections to the tree-level pole mass $m_\phi$ at the potential minimum in vacuum due to screening by the plasma}.\footnote{Here we for simplicity assume that the potential minimum is at $\varphi=0$, which is not true in many inflationary models. However, all subsequent arguments hold if we replace $\varphi$ by the deviation of the field from its value at the minimum.} 
The key point is that the same types of Feynman diagrams from which $\GG$ and $\V$ are computed also appear in the $\Phi$-self-energy and therefore determine the pole of the $\Phi$-propagator, which defines  $\W_\phi$.
Hence, oscillations with an amplitude that is large enough that non-linear terms in the equation of motion for $\varphi$ dominate automatically imply large $\varphi$-dependent corrections to the pole of the \corr{$\phi$}-propagator that violate the adiabaticity condition
\eqref{adiabaticity}.

To illustrate this, let us consider a symmetric tree level potential with a ground state at vanishing field value, which we expand as $V(\Phi) = \frac{1}{2}m_\phi^2\Phi^2 + \frac{1}{4!}\lambda_\phi\Phi^4 + \mathcal{O}[\Phi^6]$.
If we (for simplicity) place $\Phi$ in a thermal bath of temperature $T$, 
the dispersion relation reads $\disprel_\phi^2 \simeq \W_\phi^2 =  {\bf k}^2 + \mass_\phi^2 $,
while we find 
$\partial^2_\varphi\V(\varphi) = \mass_\phi^2 + 
\mathcal{O}[\varphi^6]$
, with the same correction $\mass_\phi^2 = m_\phi^2 + \lambda_\phi T^2/4! + \frac{\lambda_\phi}{2}\varphi^2 = \M_\phi^2 + \frac{\lambda_\phi}{2}\varphi^2$.\footnote{Here we have taken the high temperature limit and only considered the leading term, more accurate expressions can e.g.~be found in \cite{Curtin:2016urg,Buldgen:2020umy} and references therein.}
The quartic term in $\V(\varphi)$ dominates for $\varphi^2> \varphi_c^2$, with
\begin{equation}\label{varphicdef}
\varphi_c \equiv 12 \M_\phi^2/\lambda_\phi.
\end{equation}
When the amplitude of the oscillations greatly exceeds $\varphi_c$
we can approximate $\V(\varphi)\simeq \lambda_\phi \varphi^4/4!$.
In this regime the $\varphi$-dependent term $\frac{\lambda_\phi}{2}\varphi^2$
in $\mass_\phi^2$
is larger than the $\varphi$-independent term $\M_\phi^2$. Hence, $\mass_\phi^2$ is strongly time dependent, leading to a violation of the condition \eqref{adiabaticity}.
Neglecting dissipation, \eqref{EOM} can  be solved by\footnote{This approximation is conservative because the oscillations usually do not immediately start after the end of inflation, but there is a brief period of ``fast roll". Also neglecting this is conservative in the present context because $\varphi$ would decrease during this period.} 
\begin{eqnarray}
\varphi(t)\simeq 
\begin{cases}
\varphi_{\rm end}{\rm cn}(1/\sqrt{2},\sqrt{\lambda_\phi/6}\varphi_{\rm end} t) & \varphi_{\rm end} \gg \varphi_c\\
\varphi_{\rm end}\cos(\M_\phi t) & \varphi_{\rm end} \ll \varphi_c
\end{cases}
\end{eqnarray}
where $\varphi_{\rm end}$ is the field value at the end of inflation. 
Expanding the elliptic cosine\footnote{To be explicit, $cn(1/\sqrt{2},x)= A \sum_{n=0}^\infty c_n \cos(b_n x)$ with
$A=4.79257$, 
$c_n=\frac{e^{-\pi(n+1/2)}}{1+e^{-2\pi(n+1/2)}}$, and $b_n=\frac{\pi}{3.70815}(2n+1)$.}
one finds that 
\begin{equation}\label{varphioftcosapprox}
\varphi(t)\simeq \varphi_{\rm end} \cos(\omega t)
\end{equation} 
in both, the quadratic regime (amplitude $\ll \varphi_c$) and the quartic regime (amplitude $\gg  \varphi_c$), with 
\begin{eqnarray}\label{omegas}
\omega \simeq \sqrt{\lambda_\phi/6} \ \varphi_{\rm end} \ {\rm for} \ \varphi_{\rm end}\gg \varphi_c \quad , \quad \omega \simeq \M_\phi \ {\rm for} \ \varphi_{\rm end}\ll \varphi_c,
\end{eqnarray}
which can be inserted into $\W_\phi$. Demanding that \eqref{adiabaticity} holds at all times imposes an upper bound on $\lambda_\phi\varphi_{\rm end}^2$ that
can be written as 
\begin{eqnarray}
\lambda_\phi\varphi_{\rm end}^2 \ll \omega^2 \f(\w_\phi/\omega)
\end{eqnarray}
The most general expression for $\f(x)$
is lengthy and not illuminating.
For the present discussion it is sufficient to know that 
\begin{eqnarray}
\f(x)  
\simeq \begin{cases}
\frac{27}{2} x^4 & x > 1 \\
4 x^3 & x \ll 1
\end{cases},
\end{eqnarray}
and that $\f(1)\simeq 14.45$. 
Together with \eqref{omegas} and the knowledge that $\varphi_c \gg \M_\phi$ 
this shows that \eqref{adiabaticity} can never be fulfilled for $\varphi_{\rm end} > \varphi_c$.
In the quadratic regime $\varphi_{\rm end} < \varphi_c$, on the other hand, the adiabaticity condition translates into 
$\lambda_\phi\varphi_{\rm end}^2 \ll \M_\phi^2 \f(1) \simeq \frac{27}{2}\w_\phi^4/\M_\phi^2$, which is always fulfilled for $\varphi_{\rm end}\ll \varphi_c$ and mildly saturated for $\varphi_{\rm end}= \varphi_c$. Hence, \eqref{adiabaticity} is always violated in the quartic regime and generally fulfilled in the quadratic regime. As stated before, the reason lies in the fact that the same types of terms appear in the equation of motion for $\varphi$ and the $\Phi$-propagator.\footnote{
This connection is of course absent when it comes to the adiabaticity condition for the modes of fields $\X$ other that $\Phi$, whose effective mass $M_\X$ is not directly connected to $\V(\varphi)$  and may be smaller or larger than $\M_\phi$. Hence, $\w_\X$ may be smaller or larger than $\omega$ in both, the quadratic or quartic regimes (and even for larger elongations), depending on the vacuum mass of $\X$, its interactions, and $T$, all of which enter $M_\X$ and are part of the $\{\SM_i\}$. 
}
Hence, a  determination of $\g$ independent of the underlying particle physics model is only possible if the inflaton oscillations occur in the mildly non-linear regime with
\begin{eqnarray}\label{MildlyNonlinear}
\varphi_{\rm end} < \varphi_c \ , \ \mass_\phi \simeq \M_\phi  \ ,  \ \omega \simeq \M_\phi \ , \ \Vend\simeq \frac{1}{2}M_\phi^2 \varphi_{\rm end}^2. 
\end{eqnarray}
This conclusion will be backed up by a more detailed computation in section \ref{sec:selfinteractions}.

\subsection{Avoiding large occupation numbers in individual modes}\label{SingleModeOccupation}
Large occupation numbers 
can be avoided in two ways: Either the produced particles decay into other particles within one oscillation, or the expansion of the universe redshifts the particles' momenta out of the resonance bands. 
Since the decay rates of the produced particles necessarily depend on the $\{\SM_i\}$, $\GG$ in such scenarios necessarily depends on the $\{\SM_i\}$.\footnote{
In the adiabatic regime with $\mass_\X\simeq \M_\X$ the decay is generally slower that the oscillations because the leading processes typically require $\M_\X<\omega$, cf.~table  \ref{ThermalRates}. At the same time one can only speak of particles for $\Gamma_\X < \M_\X$, with $\Gamma_\X$ the inverse lifetime. Hence, it is impossible to achieve a decay within one oscillation, which would require $\Gamma_\X > \omega$. 
However, the time dependence of $\mass_\X$ also induces a time-dependence of $\Gamma_\X$. Assuming $\Gamma_\X \sim \SM^2\mass_\X$ and $\mass_\X\sim \SM \varphi$, cf.~appendix \ref{ApendixQFT}, it is straightforward to see that $\Gamma_\X$ can easily exceed $\omega$ when the elongation is maximal ($\varphi\sim\varphi_{\rm end}$). This is e.g.~occurs for Higgs inflation \cite{GarciaBellido:2008ab}.
}
The redshifting, on the other hand, depends on the equation of state, which is dominated by $\varphi$, and therefore in good approximation independent of the $\{\SM_i\}$. 
Hence, redshifting is the only way to reduce the particle numbers in a $\SM_i$-independent way. 
If the range $\Deltak$ of momenta $\kk = |\textbf{k}|$ over which particles are produced is narrow enough,
 large occupation numbers can be avoided by redshifting alone, and $\GG$ is a function of $\g$ only \corr{(in addition to the $\{\vv_i\}$)}. 
This is much more difficult if the particles are produced efficiently over a broad range of momenta.  
We will therefore, to be conservative, impose two conditions,
\begin{itemize}
\item[I)] that efficient particle production occurs only for a narrow range 
$\Deltak\ll\kk$
of momenta $\kk$
and 
\item[II)] that the universe expands fast enough to redshift these momenta before the occupation numbers in mode $\kk$ reach unity.  
\end{itemize}
The second condition can be quantified by estimating how many particles are being produced during the time $\Delta t$ that it takes to redshift a momentum $\kk$ out of the band
$\Deltak$.
For $\Deltak/\kk\ll1$ we can Taylor expand in $\Delta t$ to estimate
$\Deltak/\kk = 1 - a(t)/a(t + \Delta t)\simeq H(t) \Delta t$, so $\Delta t\simeq \Deltak/( \kk H)$.  
Denoting the rate at which the occupation numbers in mode $\textbf{k}$ grow by $\Gamma_k$,  
we arrive at the simple condition
\begin{eqnarray}\label{RedshiftCondition}
\frac{\Gamma_k}{H}\frac{\Deltak}{\kk} \ll 1.
\end{eqnarray}

\paragraph{Non-perturbative particle production.} The non-perturbative particle production (iv) tends to be very efficient, leading to $\mathcal{O}[1]$ occupation numbers, strong feedback, and resonant particle production within a few oscillations \cite{Traschen:1990sw}.
We shall therefore deal with the non-perturbative production first.
Using \eqref{varphioftcosapprox}, the equations for the mode functions $\X_k$ of fields that couple to $\varphi$ can be brought into the form of the usual Mathieu equation, 
 \begin{eqnarray}
\X_k''(z) + \left[
A_k - 2q\cos(2z)
\right]\X_k(z)=0\label{Mathieu}.
\end{eqnarray}
with $z\sim t\omega$ (for the moment ignoring prefactors and phase shifts).
The Mathieu equation shows instabilities that indicate resonant particle production.
The conditions under which  resonant production occurs have been studied in detail in ref.~\cite{Kofman:1997yn}. Here we use the results of this work to phrase our conditions I) and II) in terms of $q$ and $A_k$,
\begin{itemize}
\item[I)]  The width of the resonance bands of \eqref{Mathieu} scales as  $\sim q$, so the condition to avoid a broad resonance translates into $q\ll1$.\footnote{This is a conservative criterion for which we assumed that $A_k$ can be small for small momenta. If the produced particles are heavier than $\omega$, then the range of $A_k$ may be restricted to be larger than $1$, in which case the width of the bands is smaller.}
\item[II)]  In the regime $q<1$ the rate of resonant particle production can be estimated as $\sim q \omega$. The effective width of the first band is roughly $\sim q \omega/2$, 
while the momenta of the produced particles can be estimated as $\omega/2$.
Hence, \eqref{RedshiftCondition} requires $q^2\omega\ll H$. For simplicity and to be conservative we impose the condition at the beginning of reheating, which is sufficient to ensure that the condition holds during reheating in the scenarios that we consider here because $q^2$ decreases quicker with time than $H$.
\end{itemize} 

\paragraph{Elementary decays and scatterings.} The conditions 
$q\ll1$ and $q^2\omega\ll H$
ensure that the non-perturbative particle production (iv) does not trigger a parametric resonance. 
Elementary $1 \to 2$ decays (i) and
$2\to2$
annihilations (ii) of particles at rest
also lead to high occupation numbers 
because they exclusively populate narrow ranges of modes $\textbf{k}\sim \mass_\phi/2$ and $\textbf{k}\sim \mass_\phi$, respectively.\footnote{
In principle higher harmonics in the expansion of $\varphi(t)$ beyond \eqref{varphioftcosapprox} and sub-leading terms in the expansion in the couplings both give rise to processes with energies given by integer multiples of $\omega$ in the initial state, 
cf.~e.g.~Sec.~6 in \cite{Ai:2021gtg},
leading to an extended set of possible momenta for the produced particles. For the present discussion it is sufficient to consider the leading terms because they would reach large occupation numbers first.\label{FootnoteHarmonicExpansion}
} 
 Let us consider $1 \to 2$ decays of quanta that compose the condensate into $\X$-quasiparticles of mass $\mass_\X$,
which dominate at small field values.
The range $\Deltak$ over which the produced particles' momenta $\kk$ are spread is affected by two factors: The ``natural line width" 
for fixed masses $\mass_\phi$ and $\mass_\X$, 
and the spread caused by the time dependence of the effective masses $\mass_\phi$ and $\mass_\X$.
 We can approximate the former by 
 $\Gamma_{\varphi\to\X\X}$ 
 and the latter by $q\omega$,\footnote{
 The estimate is consistent with \eqref{qscaling}.
}  
where 
$\Gamma_{\varphi\to\X\X}$ 
is the rate of elementary decays.
 Here we have used the fact 
that the discussion following \eqref{adiabaticity} restricts us to the mildly non-linear regime \eqref{MildlyNonlinear}
to set $\omega \simeq \M_\phi$ and $\mass_\phi \simeq \M_\phi$.
Further recalling that 
condition I) enforces $\mass_\X^2\ll \M_\phi^2$, and that the contribution to $\Gamma_k$ from narrow resonance in the first band is also $\sim q\omega$, we can 
conclude that $\Deltak$ can be estimated by the natural line width $\Gamma_k$ whenever perturbative decay dominates over the narrow resonance. Hence,  condition II) as expressed in Eq.~\eqref{RedshiftCondition} applied to perturbative decay
reads $\Gamma_k^2/\omega\ll H$
with $\Gamma_k\simeq \Gamma_{\varphi\to\X\X}$. 
Since $\Gamma_k\leq \GG$ and $\Gamma_k \ll \omega$, this condition is always fulfilled before the moment when $H=\GG$ and reheating ends, and the perturbative decay does not trigger a parametric resonance. 
Hence, the conditions $q<1$ and $q^2\omega\ll H$ are sufficient to avoid
$\{\SM_i\}$-dependencies of the CMB observables due to
resonant particle production triggered by quantum statistical effects.

\subsection{Avoiding plasma effects}\label{ThermalBrief}
\corr{For the discussion of plasma effects we assume that the plasma of decay products can be described by a thermal bath with an effective temperature $T$.}\footnote{This is well-justified if $\omega < \Gamma_\X$, with $\Gamma_\X$ the typical relaxation rate for the most relevant modes in the plasma of decay products $\X$. In this regime elementary decays  are, however, typically forbidden because they require $\omega > \M_\X$, cf.~table \ref{ThermalRates}, while one generally expects $\Gamma_\X<\M_\X$.
For $\omega > \Gamma_\X$ the use of an effective temperature $T$ to characterise the occupation numbers in the plasma can still be justified if the $\varphi$ elongation is small enough that deviations from \eqref{MildlyNonlinear} can be treated perturbatively,
and provided that $\Gamma_\X \gg \GG$. Since the conditions I) and II)  restrict us to scenarios with $\g\ll1$ (while the decay products can have  gauge interactions with $\SM\sim\mathcal{O}[1]$) this is a reasonable assumption. A more detailed discussion can e.g.~be found in~\cite{Mazumdar:2013gya,Harigaya:2013vwa,Harigaya:2014waa,Mukaida:2015ria}.
}
The presence of the plasma can affect $\GG$ in several ways, cf.~e.g.~\cite{Hosoya:1983ke,Morikawa:1986rp,Calzetta:1989vs,Boyanovsky:1994me,Greiner:1996dx,Kolb:2003ke,Yokoyama:2005dv,Bodeker:2006ij,Drewes:2010pf,Mukaida:2012qn,Mukaida:2012bz,Drewes:2013iaa,Mukaida:2013xxa,Drewes:2014pfa,Drewes:2015eoa,Laine:2021ego,Lebedev:2021tas,Bodeker:2022ihg} and references therein. Firstly, quantum statistical effects (induced transitions and Pauli blocking) can enhance or \corr{suppress} the elementary decays (i) and annihilations  (ii) that are already present at $T=0$ in a way that depends on the time evolution of $T$, and therefore on the $\{\SM_i\}$. 
This roughly occurs when the temperature $T$ reaches $m_\phi$.\footnote{In principle thermalised particles are less dangerous in this regard than the narrow resonance bands, as a given energy density $\rho_R$ results in lower occupation numbers in each mode when it is spread over a wider range of momenta. On the other hand this also makes it more difficult to reduce the occupation numbers by redshifting once they reach unity. This is the reason why the conditions I) and II) previously formulated are not necessarily sufficient to avoid a thermally induced resonance.}
Secondly, the presence of the plasma opens up new channels of dissipation via elementary processes, such as scatterings (iii) of the quanta in the condensate $\varphi$ with quasiparticles from the bath. These typically also become relevant for $T\sim m_\phi$.
Finally, thermal corrections to the quasiparticle dispersion relations (typically of order $\SM T$, \corr{with $\SM \in \{\SM_i\}$ the relevant coupling}) strongly affect the kinematics for $T> \M_\phi/\SM$. Since the thermal correction in $\M_\phi$ is always smaller than $T$, this practically amounts to $T> m_\phi/\SM$.
We give some explicit examples to illustrate this behaviour in table \ref{ThermalRates}.

Hence, we can neglect thermal feedback on $\GG$ if $T$ remains below $m_\phi$.\footnote{One may wonder to what degree this argument can be generalised beyond the specific interactions given in table \ref{ThermalRates}. In fact this generalisation is straightforward because it only relies on kinematic considerations. Neglecting subleading corrections (cf.~footnote \ref{FootnoteHarmonicExpansion}) and still restricting ourselves to the mildly non-linear regime, $1\to2$ decays and annihilations of the quanta that constitute $\varphi$ amongst each other primarily produce particles with energies $\sim \M_\phi/2$ and $\sim \M_\phi$, respectively.}
One may therefore be tempted to impose the condition $T_{\rm max} < m_\phi$, with $T_{\rm max}$ the maximal temperature during the reheating epoch.
This condition would, however, be too conservative: 
While thermal feedback on $\GG$ generally modifies the \emph{thermal history} during reheating (which can e.g.~have an impact on the abundance of thermal relics) \cite{Drewes:2014pfa,Co:2020xaf,Ming:2021eut}, it does not necessarily mean that the CMB observables are affected by the $\{\SM_i\}$ because the CMB is not directly sensitive to the thermal history during reheating \cite{Drewes:2015coa}. 
The thermal corrections only affect the observables $(n_s,A_s,r)$ if they shift the moment $\GG=H$ and thereby modify $\Nreh$, i.e., change the \emph{expansion history}. 
However, even this overly conservative criterion turns out to be weaker than the bounds derived from the requirement $q^2\omega\ll H$ derived from condition II), as we will confirm in Sec.~\ref{sec:rangeofmeasurable}.

We denote by $\Gamma_0$ the rate of perturbative reheating at $T=0$ and by $T_\n$ the temperature at which thermal corrections become sizeable, i.e, the solution of the equation $|\GG - \Gamma_0|/\Gamma_0=c$ in $T$, with $c$ some number of order unity.
For the sake of definiteness we shall use $c=1$.
From the explicit expressions in table \ref{ThermalRates} one can see that $T_\n\sim m_\phi$, neglecting numerical prefactors of order unity that depend on the process under consideration.
For the following discussion it is useful to introduce the
would-be reheating temperature $T_{{\rm re} 0}=\sqrt{\Gamma_0 M_{pl} }(\frac{90}{\pi^2g_*})^{1/4}$ 
and the
would-be maximal temperature $T_{{\rm max} 0}\simeq 
(\Gamma_0M_{pl}/g_*)^{1/4}\Vend^{1/8}$ 
that one obtains by setting
$\GG=\Gamma_0$
in \eqref{TR} and \eqref{AveragedEoM}, respectively \cite{Giudice:2000ex}.
Thermal corrections affect the \emph{thermal history of the universe} (i.e., the time evolution of $T$) if 
$T_{{\rm max} 0}> T_\n$ \cite{Drewes:2014pfa}.
For $T_\n\sim m_\phi$ and $\Vend\sim\frac{1}{2}m_\phi^2\varphi_{\rm end}^2$ this roughly happens when $\Gamma_0/m_\phi >  g_* \frac{m_\phi}{M_{pl}}\frac{m_\phi}{\varphi_{\rm end}}$.
For the $1\to2$ decays in table \ref{ThermalRates} this translates into an upper bound on the coupling constant $\g$ which parametrically reads as
$\g\ll (g_* \epsilonvarM \epsilonvar)^{1/2}$, where the prefactor depends on the interaction. 
This constraint on $\g$  is  weaker than the constraints on $\g$ that we derive from the absence of parametric resonances in the next sections.
We can therefore conclude that thermal effects do not modify $\Nreh$ in the regime where conditions I) and II) are fulfilled. 
We present a slightly more quantitative discussion of this point in appendix \ref{ThermalFeedbackAppendix}.

\begin{table}
\begin{small}
\begin{tabular}{c c c c}
 interaction &  process &  contribution to $\GG$ \\
 \hline 
$g\Phi\chi^2$ & $\varphi\to\chi\chi$ &  
$\frac{g^2}{8\pi \M_\phi}\big(1-2\M_\chi/\M_\phi\big)^{1/2}\big(1+2f_B(\M_\phi/2)\big)\theta(\M_\phi-2\M_\chi)$
&   \cite{Boyanovsky:2004dj}\\
$\frac{h}{4}\Phi^2\chi^2$ &  $\varphi\varphi\to\chi\chi$ &  
$\frac{h^2\varphi^2}{256\pi \M_\phi}\big(1-\M_\chi/\M_\phi\big)^{1/2}\big(1+2f_B(\M_\phi)\big)\theta(\M_\phi-\M_\chi)$  
& \cite{Mukaida:2013xxa}\\
$\frac{\upalpha}{\Lambda}\Phi F_{\mu\nu}\tilde{F}^{\mu\nu}$ &  $\varphi\to\gamma\gamma$ &  
$\frac{\upalpha^2}{4\pi}\frac{\M_\phi^3}{\Lambda^2}\big(1-2M_\gamma/\M_\phi\big)^{1/2}\big(1+2f_B(\M_\phi/2)\big)\theta(\M_\phi-2M_\gamma)$  
& \cite{Carenza:2019vzg}\\
$y\Phi\bar{\psi}\psi$ &  
\begin{tabular}{c} 
$\varphi\to\psi\bar{\psi}$,\\ 
$\corr{M_\psi \simeq m_\psi}$ 
\end{tabular}
&
$\frac{y^2}{8\pi}\M_\phi\big(1-2m_\psi/\M_\phi\big)^{3/2}\big(1-2f_F(\M_\phi/2)\big)\theta(\M_\phi-2m_\psi)$  & \cite{Drewes:2013iaa}\\
 \ & 
 \begin{tabular}{c} 
$\varphi\to\psi\bar{\psi}$,\\ 
\corr{$M_\psi \gg m_\psi$}
\end{tabular}
 & 
 $\frac{y^2}{8\pi}\M_\phi\big(1-2M_\psi/\M_\phi\big)^{1/2}\big(1-2f_F(\M_\phi/2)\big)\theta(\M_\phi-2M_\psi)$ 
& 
\cite{Drewes:2013iaa}\\
\end{tabular}.
\end{small}
\caption{\label{ThermalRates}
Contributions to $\GG$ from different decay and annihilation processes in a thermal plasma involving scalars $\Phi$ and $\chi$, fermions $\psi$ with gauge interactions, and U(1) gauge bosons $\gamma$ with field strength tensor $F_{\mu\nu}$. Small letters denote vacuum masses and capital letters \corr{denote effective masses in a thermal bath (including the vacuum mass and thermal corrections)}.
$f_B(\omega)=(e^{\omega/T} - 1)^{-1}$ and $f_F(\omega)=(e^{\omega/T} + 1)^{-1}$ are the Bose-Einstein and Fermi-Dirac distributions, respectively, and $\Lambda$ is some heavy mass scale. 
Some of the rates given here have been computed for particle decays (rather than condensate decay), but we can take advantage of the fact that these rates in the regime \eqref{MildlyNonlinear} are at leading order identical for $1\to2$ decays, cf.~appendix \ref{ApendixQFT} and footnote \ref{FootnoteHarmonicExpansion}.
}
\end{table}

\section{The range of measurable inflaton couplings}\label{sec:rangeofmeasurable}

In the following we study the conditions introduced in section \ref{SecConditions} quantitatively for specific types of interactions. The parametric dependence of the restrictions on $\g$ turns out to be rather generic, indicating that these results hold for a broad range of interaction terms. 
Large occupation numbers can be avoided if $\g$ is sufficiently small, so that the transfer of energy from $\varphi$ to radiation is rather slow and the occupation numbers are reduced due to redshift before they can reach unity.\footnote{
Since the relation \eqref{GammaConstraint} constrains $\GG$ at the moment when reheating ends, it is strictly speaking not necessary to demand that the occupation numbers remain below unity at all times as long as the feedback does not enhance $\GG$ so much that it exceeds the rate of Hubble expansion $H$. An enhanced $\GG$ that remains below $H$ modifies the thermal history, but not $\Nreh$ \cite{Drewes:2014pfa,Drewes:2015coa}. 
More precisely, our results still apply as long as $\GG$ is dominated by elementary processes in the moment when $\GG=H$.
}
On the other hand, 
if $\g$ is too small, the universe fails to reheat the universe efficiently. A model-independent condition can be phrased as
\begin{itemize}
\item[III)] $\Treh$ must be larger than the temperature $T_{\rm BBN}$ needed to explain the observed abundances of light elements with  BBN.
\end{itemize}
The earliest process that directly \corr{affects} BBN is the freeze-out of the SM neutrinos, which occurs at temperatures around $1$~MeV for neutrino momenta of order of the temperature. We shall take $T_{\rm BBN} = 10$ MeV as a conservative estimate to avoid any impact on BBN. Most scenarios of baryogenesis or dark matter production require much higher temperatures, but the precise value is strongly model-dependent. The condition $\Treh > T_{\rm BBN}$ is both conservative and robust in the sense that is it model-independent and based on observation. 
Note that condition III) is qualitatively different from conditions I) and II) because it is a hard observational bound; failure to fulfil it implies that a given scenario is ruled out. Conditions I) and II), on the other hand, only restrict the range of values for which $\g$ can \corr{be} constrained from CMB data independently of the \corr{$\{\SM_i\}$}, and may be violated in nature.

In \cite{Drewes:2017fmn} it was found that the constraint \eqref{GammaConstraint} on $\GG$ can, roughly speaking, be converted into a constraint on $\g$ if 
$\g<10^{-5}$. The reheating temperature is high enough to be consistent with the observational constraint from BBN if $\g > 10^{-15}$. This leaves about ten orders of magnitude for which the inflaton coupling can be ``measured" from the CMB.
These results reported in ref.~\cite{Drewes:2017fmn} are rather independent of the specific interaction that couples the inflaton to radiation (i.e., whether $\g$ is a coupling to bosons or fermions).
In the present work we 
use simple well-known relations to 
investigate the origin of this general behaviour, which we find is not restricted to $\alpha$-attractor models.

\subsection{Self-interactions and the effective potential}\label{sec:selfinteractions}

Splitting $\Phi = \varphi + \fluct$ into the expectation value $\varphi$ and fluctuations $\fluct$, we shall first consider the production of inflaton particles $\fluct$ due to self-interactions. 
For later use, we consider general couplings of the form
\begin{equation}\label{WilsonExpansion}
\mathcal{L}\supset \wilson_{(n,m)}\Phi^n\chi^m\Lambda^{4-m-n}
\end{equation}
between $\Phi$ and another scalar field $\chi$,
with dimensionless Wilson coefficients $\wilson_{(n,m)}$ and some mass scale $\Lambda$.
For terms with mass dimension $D>4$ (with $D= n+m$ here) we shall interpret \eqref{WilsonExpansion} as an effective field theory with a cutoff $\Lambda$ that is larger than all other physical scales. In terms with $n+m<4$, on the other hand, one should not interpret $\Lambda$ as the cutoff in an effective field theory, but rather as a conveniently chosen scale to make the coupling dimensionless. We may therefore 
assume 
\begin{eqnarray}\label{EFTinterpret}
\Lambda = m_\phi \ {\rm for} \ n+m < 4 \ , \quad 
\Lambda \gg m_\phi, T, \varphi \ {\rm for} \ n+m>4 \ , \quad 
\end{eqnarray}
where the former choice is arbitrary and for convenience.
The terms in \eqref{WilsonExpansion} generate terms $\tbinom{n}{l}\wilson_{(n,m)}\varphi^l\fluct^{n-l}\chi^m\Lambda^{4-m-n}$ with $l\leq n$.
$\varphi(t)^l = [\varphi_{\rm end}\cos(m_\phi t)]^l$, 
can be decomposed into a series of terms $\propto\varphi^l\cos(\omega t)$, with $\omega$ given by integer multiples of  $m_\phi$ up to $l m_\phi$.
These operators generate time dependent masses for $\fluct$ and $\chi$, leading to particle production. 
At tree level, only the terms with 
$(m,l)=(0,n-2)$ 
and 
$(m,l)=(2,n)$ 
contribute.\footnote{
A systematic way to include quantum corrections has e.g.~been developed in \cite{Berges:2002cz}, cf.~appendix \ref{ApendixQFT}. 
}

The terms with $m=0$ and $n-l=2$ in  \eqref{WilsonExpansion} lead to tree level contributions to the effective $\fluct$ masses $\mass_\phi$ and can therefore induce self-resonances. 
The criterion 
\eqref{MildlyNonlinear} is a necessary, but not a sufficient one to avoid a self-resonance. 
The mode equation for $\fluct$ then receives contributions from all terms in \eqref{WilsonExpansion} with $n-l=2$ and $m=0$,
\begin{equation}
\ddot{\fluct}_k(t) + \left[\textbf{k}^2 + m_\phi^2 + 
\sum_n n(n-1)\wilson_{(n,0)}\varphi^{n-2}\Lambda^{4-n}
\right]\fluct_k(t)=0\label{ModeEqEta}
\end{equation}
Here we consider only renormalisable interactions, as terms with $n+m>4$ are suppressed by negative powers of $\Lambda$. This leaves us with the terms for $n=3$ and $n=4$. Assuming that only one or the other is important at a given time, 
we can use the following relations 
\begin{eqnarray}
A_k=\frac{(2\omega_\phi)^2}{\omega^2}, \quad q=2g_\phi\frac{\varphi_{\rm end}}{\omega^2}, \quad z=(t\omega+\pi)/2 \quad {\rm [ for \ the}\  \frac{g_\phi}{3!}\Phi^3 \ \rm{interaction]}\label{Akqquadratic}\\
 A_k 
= \frac{\omega_\phi^2}{\omega^2} + 2q
 \quad , \quad q=\frac{\lambda_\phi}{8}\frac{\varphi_{\rm end}^2}{\omega^2} \quad , \quad z=t\omega+\pi/2 \quad
 {\rm [ for \ the} \  \frac{\lambda_\phi}{4!}\Phi^4 \ \rm{interaction]},\label{Akqquartic}
\end{eqnarray}
where we have defined $\wilson_{(4,0)}=\lambda_\phi/4!$\corr{,} $\wilson_{(3,0)}=g_\phi/(3!\Lambda)$ \corr{and $\omega_\phi = \sqrt{\textbf{k}^2 + m_\phi^2}$}.
Using the standard relations
 \eqref{Friedmann}  
this can be translated into 
\begin{eqnarray}
\lambda_\phi \ll 8 \frac{\omega^2}{\varphi_{\rm end}^2} \ , \ 
\lambda_\phi \ll 8\sqrt{\frac{2}{3}}\frac{\Vend^{1/4}}{\varphi_{\rm end}}\frac{\omega}{\varphi_{\rm end}}\sqrt{\frac{\omega}{M_{pl}}} \ , \
g_\phi \ll \frac{6\omega^2}{
\varphi_{\rm end}} \ , \
g_\phi \ll \sqrt{24}\frac{\Vend^{1/4}}{\varphi_{\rm end}}\omega \sqrt{\frac{\omega}{M_{pl}}}.
\label{SelfInteractionConditions}
\end{eqnarray}
Using \eqref{MildlyNonlinear} 
with $m_\phi \sim M_\phi$  at the end of reheating (cf.~Sec.~\ref{ThermalBrief}),
this simplifies to 
\begin{eqnarray}\label{SelfInteractionsInTermsOfEpsilons}
\lambda_\phi\ll 8 \epsilonvar^2 \ , \quad \lambda_\phi\ll 5.5 \ \epsilonvar^{3/2}\epsilonvarM^{1/2} \ , \quad 
g_\phi/m_\phi \ll 6\epsilonvar  \ , \quad
g_\phi/m_\phi
\ll 33 \ \epsilonvar^{1/2}\epsilonvarM^{1/2}.
\end{eqnarray}
in the notation \eqref{Smallness}.
Ignoring all numerical factors, one can therefore roughly summarise \eqref{SelfInteractionsInTermsOfEpsilons} as
\begin{equation}\label{SelfInteractionSimple}
\wilson_{(4,0)}
\ll  {\rm min}(\epsilonvar^2, \ \epsilonvar^{3/2}\epsilonvarM^{1/2}) 
\ ,\quad 
\wilson_{(3,0)} \ll  {\rm min}(\epsilonvar, \ \epsilonvar^{1/2}\epsilonvarM^{1/2}) 
\end{equation}
\corr{
At this point we shall question our restriction to values $n\leq 4$, which we justified with the suppression of higher order terms in \eqref{ModeEqEta} by powers of $\varphi/\Lambda$, assuming  that $\Lambda$ must be larger than $\varphi$ in~\eqref{EFTinterpret}. 
This assumption is motivated by an 
interpretation of \eqref{WilsonExpansion} as an effective field theory during reheating \cite{Ozsoy:2017mqc}
with a cutoff $\Lambda$  that exceeds all other relevant scales.
Terms with $\Lambda \sim M_{pl}$ can certainly be expected from gravitational effects. 
An effective field theory approach with a lower cutoff can also be justified after integrating out fields with masses that exceed $m_\phi$ and the energies of particles produced during reheating, in which case $\Lambda$ should be identified with the mass of the lightest field that was integrated out.
In principle there is no reason why $\varphi$ should remain smaller than the cutoff, as $\varphi$ does not correspond to the mass of any physical particle. However, the non-perturbative mechanism (iv) can produce particles with masses much larger than $m_\phi$ \cite{Chung:1998bt}.
This is evident from the fact that the Mathieu equation \eqref{Mathieu} has resonance bands for $A_k \gg 1$.}\footnote{ 
\corrmaybe{The heaviest particles produced gravitationally roughly have a mass $m_\X$ corresponding to $H$ at the end of inflation. Using \eqref{Friedmann},
\eqref{MildlyNonlinear} and $\M_\phi\simeq m_\phi$ we can estimate the maximal $m_\X$ as $ \sim m_\phi \varphi_{\rm end}/M_{pl}$. 
For non-perturbative production from $\varphi$-oscillations we can use \eqref{adiabaticity} and footnote \ref{AdiabaticityFootnote}.
With \eqref{Akforchi} and \eqref{AkHiggsPortal} one parametrically finds that the maximal $m_\X$ scales as $\sim (g\varphi_{\rm end} m_\phi)^{1/3}$ and $\sim \varphi_{\rm end}\sqrt{h/8}$, indicating that \eqref{EFTinterpret} may be too conservative, and using a cutoff $\Lambda < \varphi_{\rm end}$ can be justified even during preheating.  
}}
\corr{Hence, when using a cutoff $\Lambda < M_{pl}$ one has to check explicitly that particle production from terms with $n>4$ does not lead to feedback effects.}

In summary, the previous considerations show that the upper bound on the inflaton self-couplings from the requirement that $\GG$ is independent of the $\{\SM_i\}$ is given by the ratios \eqref{Smallness} of $m_\phi$ and other physical scales, and that the power at which these ratios appear is determined by the power $n$ at which $\Phi$ appears in the respective term. More precisely, the power of the small parameters is given by the power at which $\varphi$ appears in $\W_{\bf k}^2$. 
These conditions appear to be consistent with what was found in a\corr{n} 
analysis for the specific case of $\alpha$-attractor models in~\cite{Lozanov:2017hjm}.

Before moving on, we shall briefly come back to the point that $\g$ can only be independent of the $\{\SM_i\}$ if $\varphi_{\rm end}$ is small enough
the oscillations happen in the mildly non-linear regime in which $\V(\varphi)$ is approximately parabolic, which we already  mentioned after \eqref{adiabaticity}.
It is straightforward to see that it is impossible to fulfil condition I) by plugging \eqref{omegas} into \eqref{SelfInteractionConditions}.\footnote{\label{AdiabaticityFootnote}
In this context it is instructive to compare the requirement $q\ll1$ from condition I) to the adiabaticity condition \eqref{adiabaticity} 
that has to be violated to efficiently produce heavy particles. For the $\wilson_{(3,0)}\Phi^3$ interaction the adiabaticity condition \eqref{adiabaticity} translates into $2q\ll A_k^{3/2}$, for quartic interaction $\lambda_\phi\Phi^4/4!$ one finds $q\ll (6A_k + 9A_k^2 - \frac{2}{3}((1+6A_k)^{3/2} -1) )^{1/2}/6$. For $A_k\ll1$ this coincides with the condition $2q\ll A_k^{3/2}$, and for $A_k\gg1$ is reproduces the commonly used condition $2q\ll A_k$.
Note that $(\omega_\phi/\omega)^2\geq 1$ in the quadratic regime while $(\omega_\phi/\omega)^2 < 1$ in the quartic regime, so that condition I) tends to be stronger than \eqref{adiabaticity} in the quadratic regime while \eqref{adiabaticity} is stronger in the quartic regime.
}

\subsection{Interactions with other scalars}\label{OtherScalarSec}

The mode equation for $\chi$  receives contributions from all terms in \eqref{WilsonExpansion} with 
\corr{$n=l$} 
and $m=2$,
\begin{equation}
\ddot{\chi}_k(t) + \left[\textbf{k}^2 + m_\chi^2 + 
\sum_n 2\wilson_{(n,2)}\varphi^{n}\Lambda^{2-n}
\right]\chi_k(t)=0\label{ModeEq2}.
\end{equation}
Comparing to \eqref{Mathieu}, we can infer that the parameter which $q$ quantifies the amplitude of the time-dependent mass and has a contribution that parametrically scales as 
\begin{equation}\label{qscaling}
q\sim \wilson_{(n,2)}\varphi^n \Lambda^{2-n}/\omega^2.
\end{equation}
Both conditions I) and II) impose upper bounds on $q$ and therefore $\wilson_{(n,2)}$. 
From \eqref{qscaling} it is clear that the power $n$ at which $\Phi$ appears in \eqref{WilsonExpansion} is decisive, and that the upper bound on the coupling $\wilson_{(n,2)}$ is stronger for larger $n$. 
We therefore classify the different interactions according to $n$, which turns out to be more relevant than e.g.~the spin of the produced particles. 

\paragraph{Scalar two body decays.}
Consider a coupling of the form $g\Phi\chi^2$ to another scalar $\chi$, where $g=\wilson_{(1,2)}\Lambda$ in the notation of \eqref{WilsonExpansion}.
The equation for $\chi$-modes in Minkowski space reads 
\begin{equation}
\ddot{\chi}_k(t) + \left[\textbf{k}^2 + m_\chi^2 + 2g\varphi(t)\right]\chi_k(t)=0\label{ModeEq}
\end{equation}
We can bring the mode equation \eqref{ModeEq} into the form \eqref{Mathieu}
with
\begin{eqnarray}\label{Akforchi}
A_k=\frac{(2\omega_\chi)^2}{\omega^2} \quad , \quad q=4\frac{g\varphi_{\rm end}}{m_\phi^2} \quad , \quad z=(t \omega+\pi)/2 \quad 
 {\rm [ for \ the} \  g\Phi\chi^2 \ \rm{interaction]},
\end{eqnarray}
\corr{with $\omega_\chi = \sqrt{\textbf{k}^2 + m_\chi^2}$}. 
With the $T\to 0$ limit of the rate for $\varphi \to \chi\chi$ decays given in table \ref{ThermalRates}
and using \eqref{MildlyNonlinear} as well as \eqref{EFTinterpret}
 the conditions I)-III) translate into 
 \begin{eqnarray} \label{ConditionsScalar}
\varphi_{\rm end} \frac{g}{m_\phi} < \frac{m_\phi}{4} 
\ 
, 
\quad
 \varphi_{\rm end} \frac{g}{m_\phi}<\sqrt{\frac{m_\phi}{
 24 M_{pl}}}\Vend^{1/4} 
 \ 
 , 
 \quad  
 \frac{g}{m_\phi} > \frac{T_{\rm BBN}}{\sqrt{m_\phi M_{pl}}}\pi \left(g_* \frac{32}{45}\right)^{1/4}.
\end{eqnarray}
\corr{W}e can simplify this to
\begin{eqnarray}\label{SimpleScalarConstraints}
\frac{g}{m_\phi} \ll \frac{1}{4}\epsilonvar \ , 
\ \frac{g}{m_\phi} \ll 0.2 \sqrt{\epsilonvar \epsilonvarM} \ , 
\  \frac{g}{m_\phi} \gg \frac{T_{\rm BBN}}{m_\phi} 3\pi \sqrt{\epsilonvarM} \left(\frac{g_*}{106.75}\right)^{1/4},
\end{eqnarray}
where $106.75$ is the value of $g_*$ in the symmetric phase of the SM and we have rounded all numerical prefactors to ratios of integers.
If we set $T_{\rm BBN}\simeq 10 \ {\rm MeV}$, $\varphi_{\rm end}\sim M_{pl}$,\footnote{This is conservative because in reality $\varphi$ is typically sub-Planckian when the oscillations commence.}  and ignore factors of order one, we \corr{obtain} the very simple condition
\begin{equation}\label{VerySimpleConstraint}
  10^{-10} \left(\frac{g_*}{106.75}\right)^{1/4} \sqrt{\frac{\rm GeV}{m_\phi}} < \frac{g}{m_\phi} < 10^{-19} \frac{m_\phi}{\rm GeV}. 
 \end{equation}
\corr{The set of inequalities \eqref{VerySimpleConstraint}}
marks the range of inflaton couplings for which reheating is driven by elementary processes. Since medium effects are negligible in this region (cf.~Sec.~\ref{ThermalBrief}), we can use the $T\to 0$ limit of the expressions given in table~\ref{ThermalRates} for $\GG$ to translate the relation \eqref{GammaConstraint} into a constraint on the inflaton coupling.
If the produced particles are massless, this in the present case yields the very simple well-known expression $\Gamma_{\varphi\to\chi\chi}=g^2/(8\pi m_\phi)$.

\paragraph{Inflaton annihilations.}
We now consider the operator $\wilson_{(2,2)}\Phi^2\chi^2\equiv\frac{h}{4}\Phi^2\chi^2$.
Comparing \eqref{ModeEq2} and  \eqref{ModeEqEta} it is straightforward to obtain the expressions for $A_k$ and $q$ by replacing $\lambda_\phi\to h$ in \eqref{Akqquartic},
\begin{eqnarray}
A_k 
= \frac{\omega_\chi^2}{\omega^2} + 2q
 \quad , \quad q=\frac{h}{8}\frac{\varphi_{\rm end}^2}{\omega^2} \quad , \quad z=t\omega+\pi/2 \quad
 {\rm [ for \ the} \  \frac{h}{4}\Phi^2\chi^2 \ \rm{interaction]},\label{AkHiggsPortal},
\end{eqnarray}
and obtain from \eqref{SelfInteractionSimple}
\begin{eqnarray}\label{SimpleScalarSmallnessesh}
h \ll {\rm min}(\epsilonvar^2, \ \epsilonvar^{3/2}\epsilonvarM^{1/2}).
\end{eqnarray}
In order to evaluate condition III) we need the annihilation rate during adiabatic harmonic oscillations  
from table~\ref{ThermalRates}
for which we take $\Gamma_{\varphi\varphi\to\chi\chi} \simeq \frac{h^2 \varphi^2}{256\pi m_\varphi}$ from table \ref{ThermalRates}. 
This rate can intuitively be interpreted as an annihilation of two $\Phi$-quanta into two $\chi$-quanta. 
Determining $T_R$ is a bit more tricky because $\GG\propto\varphi^2$, while $H\propto|\varphi|$.
As a result, the ratio $\GG/H$ (averaged over a few oscillations) decreases with time.
For $\varphi_{\rm end} > 256\sqrt{2}\pi m_\varphi^2/(3 h^2 M_{pl})$ this ratio is larger than one already when the oscillations commence, for smaller $\varphi_{\rm end}$ it remains below one at all times, and the universe is not reheated. The requirement to reheat the universe through inflaton annihilations can be translated into the condition on $h > 16\times 2^{1/4} \sqrt{\pi/3}\sqrt{\epsilonvar\epsilonvarM}$. 
Comparing this to \eqref{SimpleScalarSmallnessesh} shows that 
it is not possible to reheat the universe with elementary annihilations (as it was e.g.~assumed in \cite{Almeida:2018oid}):
one either encounters a resonance, or the universe is never reheated.

 Before moving on, we should add that several details have been wiped under the carpet here.
We implicitly assumed an averaging over a few oscillations when comparing $\GG$ to $H$ without specifying how exactly this averaging should be done. This may be important because the $\varphi$-dependence of $\GG$ introduces a non-linearity in \eqref{EOM}, so that the motion of $\varphi$ is in general not that of a simple damped harmonic oscillator even if we set $\V=\frac{1}{2}m_\phi^2\varphi^2$. As a result the relation between the rates of elementary processes and the damping of $\varphi$ can be more complicated than in the linear regime \cite{Ai:2021gtg}.

\subsection{Two body decays into fermions}\label{YukawaSec}
 It is usually assumed \corr{that} reheating primarily produces bosons because quantum statistical effects lead to an enhancement of the rate, while they suppress the production of fermions due to Pauli's principle. 
However, fermions can also be produced, and they can also experience resonant enhancement \cite{Greene:1998nh}.\footnote{This can have interesting implications even if the fermion production is sub-dominant compared to that of bosons, e.g.~for leptogenesis or gravitino production \cite{Giudice:1999fb,Greene:2000ew}.}
For the present discussion the fermion production is only relevant if it dominates the reheating process and sets $\Nreh$. 
This requires that the coupling to fermions is considerably larger than that to light bosons.
We here consider the production of a Dirac fermion $\psi$ of mass $m_\psi$ through a Yukawa coupling $y\Phi\bar{\psi}\psi$.
A resonance parameter can be defined as $q=y^2\frac{\varphi^2}{\omega^2}$, which yields $q\sim \frac{y^2\varphi^2}{m_\phi^2}$ if the potential is approximately quadratic \cite{Greene:1998nh,Greene:2000ew}.
The underlying mechanism for the resonant production of fermions is quite different from bosons. 
Hence, we cannot use condition II) here. In spite of the differences in the underlying microphysics, it turns out that the condition $q\ll1$ still provides a good criterion to determine the largest coupling where reheating is perturbative \cite{Greene:1998nh}.\footnote{
It should be noted that the treatment used in \cite{Greene:1998nh} underestimates the fermion production in the presence of large bosonic occupation numbers \cite{Berges:2010zv}. 
This is, however, not relevant in the present discussion because in that case the coupling to bosons would set $\Nreh$.
}
The lower bound on $y$ can again be obtained from condition III). Comparing the perturbative decay rate
$\Gamma_{\varphi\to\psi\psi}=\frac{y^2}{8\pi} m_\phi$ to $\Gamma_{\varphi\to\chi\chi}$ reveals that we can literally apply the lower bound in conditions \eqref{ConditionsScalar}-\eqref{VerySimpleConstraint} with the replacement 
$\frac{g}{m_\phi}\to y$.
We finally arrive at the range 
\begin{equation}\label{FermionSimplifiedConditions}
 \frac{T_{\rm BBN}}{m_\phi}  3\pi \sqrt{\epsilonvarM} \left(\frac{g_*}{106.75}\right)^{1/4}\ll  y \ll \frac{1}{4}\epsilonvar \end{equation}

\subsection{Gauge boson production from axion-like coupling}\label{ALPSec}
Let us now consider axion-like couplings $\upalpha \Phi \Lambda^{-1} F_{\mu\nu}\tilde{F}^{\mu\nu}$ to the field strength tensor $F_{\mu\nu}$ of vector bosons.\footnote{
Though axions may not be needed to solve the strong $CP$ problem \cite{Ai:2020mzh}, axion-like couplings appear in many theories beyond the Standard Model.
} 
We assume that the gauge bosons are massless for $\varphi=0$, which is justified in the symmetric phase of gauge theories, and consider an Abelian gauge theory.
This situation is qualitatively different from the scalar interaction because the vertex is momentum dependent, and the time dependent mass term of the gauge fields have a non-trial Lorentz structure. 
The Mathieu equation for the circular polarisations of the photon field reads \cite{Carenza:2019vzg}
\begin{equation}
\left(\partial_t^2 + {\rm k}^2 \mp 4\upalpha \frac{{\rm k} \ m_\phi}{\Lambda} \varphi \right)
a_{k \pm} = 0
\end{equation}
Comparison to \eqref{ModeEq} gives
$A_k = (2 {\rm k}/\omega)^2$ 
with ${\rm k}=|{\bf k}|$
and 
$q=8\upalpha\frac{\rm k}{\omega} \frac{\varphi}{\Lambda}\frac{m_\phi}{\omega}$.
It is immediately clear that condition I) is always violated for large momenta. On the other hand this does not necessarily imply that non-perturbative particle production dominates because for sufficiently small $\upalpha$ almost all modes remain adiabatic.\footnote{In addition the photon dispersion relations would also be modified by finite temperature/density effects, and the plasmon mass can help avoiding a resonance. However, this effect of course depends on the underlying particle physics model and the $\{\SM_i\}$.}

Here we restrict ourselves to a  simple estimate.
Both, the perturbative decay and a narrow resonance produce particles with energy $m_\phi/2$, so that we can fix the momentum to this value and use \eqref{Akqquadratic} with the replacements $\omega_\phi\to m_\phi/2$ and 
$g\to \upalpha m_\phi^2/\Lambda$ in $A_k$ and $q$.
We obtain the upper bounds on $\upalpha$ from conditions I) and II) by making this replacement in  \eqref{ConditionsScalar}-\eqref{VerySimpleConstraint}.
To obtain the lower bound from BBN, we consider the elementary decay rate into vector bosons from table \ref{ThermalRates}, $\Gamma_{\varphi\to \gamma\gamma}=\frac{\alpha^2}{4\pi}\frac{m_\phi^3}{\Lambda^2}$.
Comparison with 
$\Gamma_{\varphi\to\chi\chi}=\frac{g^2}{8\pi m_\phi}$
reveals that the 
lower bound can be obtained 
by making the same replacement $g\to \upalpha m_\phi^2/\Lambda$ 
in $\GG$ and multiplying the total rate with a factor $2$ to account for the two photon polarisations in the final state, leading to
\begin{eqnarray}\label{SimpleAxionConstraints}
\upalpha \frac{m_\phi}{\Lambda} \ll \frac{1}{4}\epsilonvar \ , 
\ \upalpha \frac{m_\phi}{\Lambda} \ll 0.2 \sqrt{\epsilonvar \epsilonvarM} \ , 
\  \upalpha \frac{m_\phi}{\Lambda}  \gg 2 \frac{T_{\rm BBN}}{m_\phi}  \sqrt{\epsilonvarM} \left(\frac{g_*}{106.75}\right)^{1/4}.
\end{eqnarray}

\section{Discussion}

For the interactions in table~\ref{ThermalRates}
that we studied 
we find that the upper bounds 
\eqref{SelfInteractionsInTermsOfEpsilons}, 
\eqref{SimpleScalarConstraints}, \eqref{SimpleScalarSmallnessesh}, and  \eqref{SimpleAxionConstraints} on the dimensionless inflaton coupling $\g$ can be summarised in terms of the ratios \eqref{Smallness} as
\begin{eqnarray}\label{GeneralScaling}
\g \ll \epsilonvar^{n - 1/2}
{\rm min}(
\sqrt{\epsilonvarM}
,
\sqrt{\epsilonvar}
)
\left(\frac{m_\phi}{\Lambda}\right)^{4-D}, \ 
\vv_i \ll \epsilonvar^{n - 5/2}
{\rm min}(
\sqrt{\epsilonvarM}
,
\sqrt{\epsilonvar}
)
\left(\frac{m_\phi}{\Lambda}\right)^{4-D},
\end{eqnarray}
 with $D$ the mass dimension of the operator and $n$ the power at which $\Phi$ appears. 
In scalar interactions of the form \eqref{WilsonExpansion} tree-level contributions to the mass 
that can trigger a parametric resonance 
come from operators with
$D=n$ for self-interactions $\vv_i$ and $D=n+2$ for couplings $\g$ to other scalars.
The most strongly constrained interactions are the self-interactions  
and couplings to other fields with $n=2$, as for larger values of $n$ the negative powers of the small quantity $m_\phi/\Lambda$ on the RHS of \eqref{GeneralScaling} soften the constraint. 
For a Yukawa coupling to fermions we find that only the first of the two conditions in \eqref{GeneralScaling} applies, cf.~\eqref{FermionSimplifiedConditions}.
Based on the observation that the time-dependent contribution to the squared effective fermion mass scales as $\sim (\g \varphi^n)^2$ one can expect constraints of the form $\g\ll \epsilonvar^n (m_\phi/\Lambda)^{1-n}$ for fermions. This scaling is consistent with \eqref{FermionSimplifiedConditions}.

There is also a lower bound on $\g$ from condition III), i.e., the requirement to reheat the universe, which for the terms with $n=1$ in \eqref{SimpleScalarConstraints}, \eqref{FermionSimplifiedConditions} and \eqref{SimpleAxionConstraints} 
roughly reads
\begin{eqnarray}\label{Kmillominomat}
\g   \gg \frac{T_{\rm BBN}}{m_\phi}  \sqrt{\epsilonvarM}\left(\frac{m_\phi}{\Lambda}\right)^{4-D}
\sqrt{\#}g_*^{1/4},
\end{eqnarray}
where we have parameterised $\GG = \g^2 m_\phi /{\#}$ with $\#$ a numerical factor, cf.~table \ref{ThermalRates}.
Ignoring numerical prefactors, this leaves a window of values for 
$\sqrt{\epsilonvarM} T_{\rm BBN}/m_\phi
\ll 
\g (m_\phi/\Lambda)^{D-4}
\ll 
{\rm min}(\epsilonvar^n,\sqrt{\epsilonvarM}\epsilonvar^{n - 1/2})$
that can span several orders of magnitude if $m_\phi$ is large enough, and closes when $m_\phi$ violates the lower bound
\begin{eqnarray}\label{MassEffect}
m_\phi > \varphi_{\rm end} \left(
\frac{T_{\rm BBN}}{\varphi_{\rm end}}
{\rm max}\left( 1,
\sqrt{\varphi_{\rm end}/M_{pl}}
\right)
\right)^{1/(n + 1/2)},
\end{eqnarray}
which roughly reads 
$m_\phi > 10^5$ GeV for  $n=1$ and $\varphi_{\rm end}\sim M_{pl}$, 
and approaches $M_{pl}$ for larger $n$.

The constraints  
\eqref{GeneralScaling}-\eqref{MassEffect} mark the range of coupling constants and inflaton masses for which it is in principle possible to constrain the inflaton coupling $\g$ in a model-independent way. They do not say anything about how accurately one can measure $\g$ in practice. 
To make a simple estimate, we assume a model of inflation where $\V(\varphi)$ has a flat plateau shape, as this class of models is amongst those presently preferred by data \cite{BICEP2:2018kqh,Planck:2018vyg}. 
Let $\Minf$ be the scale of inflation, i.e., $\Vend \sim \Minf^4$. Then \eqref{MildlyNonlinear} implies that $m_\phi \sim \Minf^2/\varphi_{\rm end}$. Here we have assumed that $\Vend\simeq \V(\varphi_k)$ for a plateau-like potential, with $\varphi_k$
the value of $\varphi$ at the Hubble crossing of the pivot mode $k$ used in the CMB data analysis.
Using \eqref{H_k} and assuming $\varphi_{\rm end} \sim M_{pl}$ we can estimate
\begin{eqnarray}\label{InflatonMass}
m_\phi \sim  M_{pl} \sqrt{\frac{3\pi^2}{2} r A_s} \ , \quad
M \sim M_{pl} \left(\frac{3\pi^2}{2} A_s r\right)^{1/4}
\end{eqnarray}
and express \eqref{GeneralScaling} in terms of observable quantities. For $n=1$ and renormalisable interactions this yields 
\begin{eqnarray}\label{YippieYaYaySchweinebacke}
\g \ll \sqrt{\frac{3\pi^2}{2} r A_s}.
\end{eqnarray}
If we assume that reheating is driven by elementary decays in vacuum (consistent with what we found in section \ref{SecConditions}), we can use standard formula \eqref{TR} for the reheating temperature to obtain
\begin{eqnarray}\label{BBNsimple}
\g > g_*^{1/4} \frac{T_{\rm BBN}}{\sqrt{m_\phi M_{pl}}}\sqrt{\#}
\simeq
\frac{T_{\rm BBN}}{M_{pl}} \left(\frac{g_*}{A_s r}\right)^{1/4}\sqrt{\#}.
\end{eqnarray}
Assuming that the mild dependence on $g_*$ can be neglected, the only unknown in \eqref{YippieYaYaySchweinebacke} and \eqref{BBNsimple} is the scalar-to-tensor ratio $r$.
Plugging in the upper bound $r=0.06$ from \cite{Planck:2018vyg} into \eqref{YippieYaYaySchweinebacke} yields $\g \ll 4\times 10^{-5}$. This is about an order of magnitude larger than the electron Yukawa coupling in the SM. 
From \eqref{BBNsimple} we find $\g>10^{-17}$ for $\#=8\pi$ and the SM value of $g_*$. This implies that the current uncertainty in $\g$ extends over $12$ orders of magnitude. 
For an inflaton mass near the bound \eqref{MassEffect} 
feedback effects can only be avoided if $\Treh\sim T_{\rm BBN}$, which requires a very small value of $\g$ at the lower end of this window. 
For larger values of $m_\phi$ the maximal reheating temperature \eqref{TR} that can be achieved while respecting \eqref{YippieYaYaySchweinebacke} can be estimated as $\Treh < g_*^{-1/4}\sqrt{m_\phi^3/(\varphi_{\rm end}\#)}\sim m_\phi \sqrt{\epsilonvar}g_*^{-1/4}$.

We can now use  \eqref{YippieYaYaySchweinebacke} and \eqref{BBNsimple} to estimate the accuracy $\delta r$ with which $r$ will have to be measured in order to determine the order of magnitude of the inflaton coupling $\g$. 
Let us for a moment suppose that we have fixed all parameters $\{\vv_i\}$ in the potential $\V(\varphi)$.
From \eqref{GammaConstraint} and table~\ref{ThermalRates} we can see that
$\g^2\propto e^{-\frac{3}{2} \Nreh}$. 
Further
using \eqref{Nre}, \eqref{Nk}\footnote{Note that  the integral in \eqref{Nk} is dominated by the upper limit $\varphi_k$, where the integrand can be written as $M_{pl} \sqrt{8/r}$.} and \eqref{TakaTukaUltras} we can express the RHS in terms of observables.
The resulting expression for $\log\g$ is in principle very complicated, but  
within the observationally allowed range, one may linearly expand $\log\g$ as a function of $r$ around some value $\bar{r}$ that is typical for the given model, i.e., $\log\g \propto r-\bar{r}$ within some interval around $\bar{r}$.
For $r>10^{-7}$ the upper and lower bounds \eqref{YippieYaYaySchweinebacke} and \eqref{BBNsimple} leave a range of at least ten orders of magnitude for $\g$. 
At the same time $r$ can at most change by an amount that is smaller than the current upper bound on its value, i.e.,~$\Delta r < 6\times 10^{-2}$ \cite{Planck:2018vyg}
or less \cite{BICEP:2021xfz}.\footnote{
While it has been shown in~\cite{Drewes:2017fmn} 
for the case of $\alpha$-attractor models
that an approximately linear relationship between $\log\g$ and the deviation in $n_s$ holds across a range of values for $n_s$ that is comparable to the current observational error bar on $n_s$,
the range of validity for the linear approximation in $r-\bar{r}$ is typically smaller than the observationally allowed range of values for $r$ (note that $r$ and $n_s$ are not independent when all $\{\vv_i\}$ are fixed, and within the observational limits, one can often establish a unique relation between them). Hence, in specific models there can be considerable deviations from the rough estimate \eqref{rRough}, necessitating more detailed studies in specific models \cite{Drewes:2022nhu}.
}
Hence, the observational resolution  $\delta r$ must be less than a tenth of $\Delta r$ if one wants to constrain 
the order of magnitude of $\g$. This very rough estimate leads to the requirement\footnote{
Note that the dependence of  $r$ on $\g$ is stronger for larger  values of $r$, cf.~e.g.~\cite{Drewes:2017fmn} for an explicit example, which would facilitate the measurement of $\g$.  
} 
\begin{eqnarray}\label{rRough}
\delta r \sim 10^{-3}.
\end{eqnarray}
An accuracy of a few times $10^{-3}$ can be achieved by several future CMB observatories, including the
Simmons Observatory \cite{SimonsObservatory:2018koc}, 
South Pole Observatory \cite{Moncelsi:2020ppj},
LiteBIRD \cite{Sugai:2020pjw},
 CMB S4 \cite{CMB-S4:2020lpa}, and PICO \cite{NASAPICO:2019thw}. 
Hence, we can conclude that next generation CMB experiments may be able to pin down the order of magnitude of the inflaton coupling.

We emphasise that the estimate \eqref{rRough} assumes that all other uncertainties can be neglected, which is possible if all parameters in $\V(\varphi)$ are either predicted by theory or determined from other observables. While $\Minf$ can be constrained well from $A_s$, the current error bar on $n_s$ is too large to simultaneously fit $\g$ and additional parameters in $\V(\varphi)$ in the family of models considered in \cite{Drewes:2017fmn,Drewes:2022nhu}.
This situation may change in the future. For instance, combining CMB observations with data from the 
EUCLID satellite can reduce the error bar on $n_s$ to 
$\sigma_{n_s} = 0.00085$,
with potential further improvement when adding data from the Square Kilometre Array 
\cite{Sprenger:2018tdb}. 
The running of the scalar spectral index $n_s$ is another potential probe \cite{Adshead:2010mc} that can provide information on reheating \cite{Martin:2016oyk}, it can be particularly important if $r$ is small \cite{Easther:2021eje}. Future CMB observations will be sensitive to the running of $n_s$ \cite{Munoz:2016owz}, combining this with data from optical,
infrared and 21cm surveys 
can improve the sensitivity considerably \cite{Mao:2008ug,Kohri:2013mxa,Munoz:2016owz}. 
Another potentially interesting observable are non-Gaussianities, which can be targeted by a number of probes including CMB observations, galaxy surveys and 21cm observations  \cite{Renaux-Petel:2015bja,Sekiguchi:2018kqe,Achucarro:2022qrl}.
A quantitative study of 
the perspectives to constrain the inflaton coupling from 
these and other potential observables goes beyond the scope of the current work, which focuses on the parameters $(A_s,n_s,r)$ that are already constrained by data. In view of the expected progress on the observational side in the next decade \cite{Achucarro:2022qrl,Komatsu:2022nvu} our results strongly motivate follow-up work to identify the most promising observables. 

\section{Conclusions}

We studied the conditions under which a model-independent relation between the inflaton coupling $\g$ and the CMB parameters $(A_s, n_s, r)$ can be established, so that the microphysical parameter $\g$ can be ``measured" in the CMB.
Model-independence here refers to the details of a specific particle physics model in which a given 
model of inflation can be embedded.
We work under the assumptions that an effective single-field description 
holds during both the inflationary and reheating epochs, 
 that reheating occurs during oscillations of the inflaton condensate $\varphi$ around the minimum of its effective potential $\V(\varphi)$,
 and that radiative corrections to $\V(\varphi)$ either remain small 
 or are dominated by inflaton self-interactions 
 during both inflation and reheating.
We further assume a standard thermal history after reheating, and that the tensor modes observed in the CMB are either dominated by perturbations generated during inflation or secondary sources can be identified and subtracted. Finally we assume that the mild dependencies on other particle physics parameters summarised in section \ref{ParameterDependencies} are negligible. This e.g.~means that the number of new particles should be small enough that the dependencies $\sim g_*^{1/4}$ can be neglected.

Under these assumptions the most severe restriction comes from the necessity
to avoid feedback effects during reheating, which in general introduce a dependence of $(A_s, n_s, r)$ on a potentially large number of microphysical parameters $\{\SM_i\}$ in the underlying particle physics model, making it practically impossible to constrain any individual parameter. 
Feedback effects and dependencies on additional parameters $\{\SM_i\}$ can be avoided if conditions I) and II) outlined in \corr{section \ref{SecConditions}} are fulfilled.
These conditions can be translated into restrictions on the inflaton's self-couplings $\{\vv_i\}$ as well as its coupling $\g$ to other fields $\X$, which can generally be expressed in terms of the ratios  $\epsilonvar=m_\phi/\varphi_{\rm end}$
and $\epsilonvarM=m_\phi/M_{pl}$ defined in \eqref{Smallness}.
 \corr{They} necessarily restrict the possibility to measure $\g$ to the mildly non-linear regime where deviations from \eqref{MildlyNonlinear} can be treated perturbatively, and impose  upper bounds \eqref{GeneralScaling} on the coupling constants.
For renormalisable interactions in table~\ref{ThermalRates} 
that are linear in $\varphi$ and in plateau-like inflation models we can use \eqref{YippieYaYaySchweinebacke} to estimate the numerical value of the upper bound to be about an order of magnitude larger than the electron Yukawa coupling in the SM. 
At the same time the requirement to successfully reheat the universe leads to the lower bound \eqref{Kmillominomat} on $\g$. For  values of $r$ near the current upper bound from observation the range of $\g$  allowed by \eqref{Kmillominomat} and \eqref{YippieYaYaySchweinebacke} spans about $12$ orders of magnitude.  
This window becomes smaller for lower $m_\phi$ and closes for $m_\phi \sim 10^5$ GeV for the interactions considered here. This unfortunately suggests that it will not be possible in the foreseeable future to measure $\g$ independently in collider experiments and in the CMB, though it should be kept in mind that this is a rough estimate. 

These conditions permit to identify the range of values for the inflaton coupling for which it in principle can be constrained by CMB observations within a given model of inflation. The practical feasibility of this measurement depends on the model of inflation. 
The rough estimate \eqref{rRough} suggests that next generation CMB experiments will be able to constrain the order of magnitude of the coupling constant by measuring the scalar-to-tensor ratio $r$. Further improvement will be possible when adding data from cosmological surveys, in particular with EUCLID and SKA. 
Even a rough measurement would \corr{mark} a significant achievement from the viewpoint of both cosmology and particle physics. It provides an indirect probe of a microphysical parameter that most likely can never be measured directly in the laboratory, but shaped the evolution of the cosmos by setting the stage for the hot big bang. Further, it can provide one of the very few observational hints on how a given model of inflation can be embedded in a more fundamental theory of nature.

\section*{Acknowledgments.}

I would like to thank Gilles Buldgen, Drazen Glavan, Jin U Kang, Lei Ming, and Vincent Vennin for their helpful comments on the draft and Christophe Ringeval for helpful discussions. I also thank my grandmother Erika Drewes and my mother Roswitha Drewes for hosting me in their homes and helping us with childcare for a few weeks during the pandemic while I was working on this article.

\begin{appendix}
\section{Relation between $\Nreh$ and observables}\label{RelationToObservables}

In this appendix we briefly review the relationship between the RHS of \eqref{GammaConstraint} and observables. 
We consider single field inflation models  in which the universe is reheated by coherent oscillations of $\varphi$ around he minimum of its effective potential $\V(\varphi)$. 
In this case the evolution of $\varphi$ is governed by the equation of motion \eqref{EOM}, c.f.~appendix \ref{EOMvarphi}.
Inflation ends when the universe stops accelerating, which happens when the equation of state exceeds $w>-1/3$. 
This roughly corresponds to the moment when $\epsilon$ exceeds unity, with
\begin{equation}
    \epsilon=\frac{M^2_{pl}}{2}\left(\frac{\partial_\varphi \V}{\V}\right)^2\ , \quad 
    \eta=M^2_{pl}\frac{\partial^2_\varphi \V}{\V},
\label{slowrollpara}
\end{equation}
More precisely, we define the \emph{reheating epoch} 
as the time between the moment when 
$\epsilon|_{\varphi_{\rm end}}=1$
and the moment when the energy density of radiation $\rho_R$ exceeds the energy density $\rho_\varphi\simeq \dot{\varphi}^2/2 + \V $ of the condensate $\varphi$, i.e., the universe becomes radiation dominated ($w=1/3$).
The latter moment  roughly coincides with the moment when $\GG=H$.
We refer to the duration of the reheating epoch in \corr{terms} of $e$-folds as $\Nreh=\ln(a_{\rm re}/a_{\rm end})$.

The redshifting relation $\rho(N)=\rhoend\exp[-3\int_0^N(1+w(N'))dN']$ 
yields the second equality in \eqref{rhoreeq}
and
permits to express the Friedmann equation \corr{\eqref{Friedmann}} as
\begin{equation}\label{FriedmannNre}
    H^2=\frac{\rho_{\rm end}}{3M^2_{pl}}e^{-3N_{\rm re}(1+\bar{w}_{\rm re})}.
\end{equation}
 and the reheating temperature as
\begin{equation}\label{Tre}
	\Treh=\exp\left[-\frac{3(1+\bar{w}_{\rm re})}{4}N_{\rm re}\right]\left(\frac{40 \Vend}{g_*\pi^2}\right)^{1/4},~
\end{equation}
where $\Vend$ is the potential at the end of inflation, and we have used  
\eqref{Friedmann}. 
Since the energy density during reheating is (by definition) dominated by $\V(\varphi)$, $\wrehbar$ only depends on the $\{\vv_i\}$. In particular, for a power law potential $\V(\varphi)\propto (\varphi/M_{pl})^p$ one finds  $\wrehbar=(p-2)/(p+2)$ \cite{Turner:1983he}.
In the regime \eqref{MildlyNonlinear} permitted by conditions I) and II), this practically implies $\wrehbar=0$.
Hence, leaving aside the mild dependence on $g_*$ (which  could be avoided if we were to consider $\rhoreh$ instead of $\Treh$),
$\Nreh$ is the only \corr{entity}
on the RHS of \eqref{Tre} that is not determined by fixing the parameters \corr{$\{\vv_i\}$} in $\V $.
Hence, determining $\Treh$ boils down to establishing a relation between $\Nreh$ and $(n_s, A_s,r)$.
A detailed derivation can be found in \cite{Ueno:2016dim} and has been adapted to our notation in \cite{Drewes:2017fmn}, the result reads 
\begin{equation}
\label{Nre}
    N_{\rm re}=\frac{4}{3\Bar{w}_{\rm re}-1}\left[N_k+\ln\left(\frac{k}{a_0 T_0}\right)+\frac{1}{4}\ln\left(\frac{40}{\pi^2g_*}\right)+\frac{1}{3}\ln\left(\frac{11g_{s*}}{43}\right)-\frac{1}{2}\ln\left(\frac{\pi^2M^2_{ pl}r A_s}{2\sqrt{\Vend}}\right)\right].
\end{equation}
Here $N_k$  is the  number of $e$-folds between
the horizon crossing of a perturbation with wave number $k$ and the end of inflation,
\begin{equation}\label{Nk}
    N_k=\ln\left(\frac{a_{\rm end}}{a_k}\right)=\int_{\varphi_k}^{\varphi_{\rm end}}\frac{H d\varphi}{\dot{\varphi}}
    \approx\frac{1}{M^2_{pl}}\int_{\varphi_{\rm end}}^{\varphi_k}d\varphi\frac{\V }{\partial_\varphi \V }~.
\end{equation}
Here $a_0$ and $T_0=2.725~{\rm K}$ are the scale factor and the temperature of the CMB at the present time, respectively.
The subscript notation $H_k, \varphi_k, \epsilon_k, \corr{\varphi_k}$  indicates the value of the quantities  $H, \varphi, \epsilon, \corr{\varphi}$ at the moment when the pivot-scale $k$ crosses the horizon.
$\varphi_k$ can be expressed in terms of the spectral index $n_s$ and the tensor-to-scalar ratio $r$ by the relation 
\begin{equation}
    n_s=1-6\epsilon_k+2\eta_k~, \quad r=16\epsilon_k.
\label{nANDr}    
\end{equation}
In the slow roll approximation, we find 
\begin{equation}
\label{H_k}
	H^2_k=\frac{\V(\varphi_k)}{3M_{pl}^2}~
		=\pi^2 M_{pl}^2\frac{r A_s}{2}
\end{equation}
with $A_s=10^{-10}e^{3.043}$ \cite{Planck:2018vyg} the amplitude of the scalar perturbations from the CMB.
$\Treh$ can be expressed in terms of the observables $(n_s, A_s,r)$
by plugging \eqref{Nre} with \eqref{Nk} into \eqref{Tre}. 
 $\varphi_k$ is fixed by solving \eqref{nANDr} for $\varphi_k$, and $\Vend$ and $\varphi_{\rm end}$ can be determined by solving $\epsilon=1$ for $\varphi$.
From \eqref{nANDr} we obtain
\begin{equation}
    \epsilon_k=\frac{r}{16}~,\quad \eta_k=\frac{n_s-1+3r/8}{2},
\end{equation}
from which we find
\begin{eqnarray}\label{TakaTukaUltras}
&&\frac{\partial_\varphi \V(\varphi)}{\V(\varphi)}\Bigl|_{\varphi_k}=\sqrt{\frac{r}{8M_{pl}^2}} \ , \quad
\frac{\partial^2_\varphi \V(\varphi)}{\V(\varphi)}\Bigl|_{\varphi_k}=\frac{n_s-1+3r/8}{2M_{pl}^2}\corr{,}
\end{eqnarray}
using \eqref{slowrollpara}.
Together with \eqref{H_k} this yields three equations that can be used to relate the effective potential and its derivatives to the observables $(n_s,A_s,r)$. This is sufficient to express $\wrehbar$ and $\Nreh$ in \eqref{Nre} in terms of observables, which is all that is needed to determine the RHS of \eqref{GammaConstraint}.

\section{A slightly more detailed discussion of thermal feedback}\label{ThermalFeedbackAppendix}

In this appendix we briefly return to the question under what circumstances the modifications of the rate for elementary processes caused by the presence of a plasma of inflaton decay products can modify $\Nreh$.
We confirm the conclusion drawn in \corr{section} \ref{ThermalBrief} that such effects are negligible in the regimes where conditions I) and II) are fulfilled, and therefore do not impose any additional constraints on $\g$.

Assuming that the inflaton performs many oscillations before decaying, we can solve the coupled equations
$\dot{\rho}_\varphi+3H\rho_\varphi+\GG \rho_\varphi =0 $ and 
$\dot{\rho}_R+4H\rho_R-\GG \rho_\varphi=0$ 
for the time averaged inflaton energy density $\rho_\varphi$
and the radiation density $\rho_R$ defined in \eqref{TR}.
It is convenient to introduce the dimensionless \corr{quantities} $\Upphi\equiv\rho_\varphi a^3/m_\phi$, $R\equiv\rho_R a^4$ and $x\equiv a m_\phi = a/a_{\rm end}$ and rewrite the equations as  
\begin{eqnarray}
    \frac{d\Upphi}{dx}=-\frac{\GG}{H x}\Upphi \ , \quad \
    \frac{d R}{dx}=\frac{\GG}{H }\Upphi \label{AveragedEoM}
\end{eqnarray}
with
\begin{eqnarray}
H=
3^{-1/2}
\frac{m_\phi^2}{M_{pl}}\left(\frac{R}{x^4}+\frac{\Upphi}{x^3}\right)^{1/2} \ , \quad \ 
T=\frac{m_\phi}{x}\left(\frac{30}{\pi^2g_*}R\right)^{1/4}.\label{TandHdefs}
\end{eqnarray}

Let us first consider the case where thermal effects suppress $\GG$, as e.g.~for the decay into fermions in table \ref{ThermalRates}.
We can then use 
$T_\n > T_{{\rm re} 0}$
as a conservative criterion to avoid feedback. This translates into the bound $\Gamma_0/m_\phi < \sqrt{g_*}m_\phi/M_{pl}$. Assuming $\Gamma_0 \propto \g^2 m_\phi$ this yields $\g \ll \epsilonvarM^{1/2} g_*^{1/4}$, which is much weaker than the constraints on $\g$ that we derive from the absence of parametric resonances in \eqref{FermionSimplifiedConditions}.

To discuss the case where thermal effects enhance $\GG$ we consider the Taylor expansion
$\GG = \sum_i \Gamma_i \ (T/m_\phi)^i$. 
If $\GG$ as a function of $T$ is a monomial, the set of equations \eqref{AveragedEoM}-\eqref{TandHdefs} can be solved analytically \cite{Drewes:2014pfa,Drewes:2015coa,Co:2020xaf,Ming:2021eut}. 
We can use this fact to obtain analytic solutions for an arbitrary dependence of $\GG$ on $T$ by locally approximating the functional dependence by a monomial power law and matching the solutions \cite{Drewes:2014pfa}, provided that one term in the Taylor expansion dominates at any given temperature. For the present purpose it is sufficient to consider the parameterisation 
\begin{equation}\label{nexpansion}
    \GG = \Gamma_0 + \Gamma_\n \  (T/m_\phi)^\n,
\end{equation}
which yields $T_\n=m_\phi (\Gamma_0/\Gamma_\n)^{1/\n}$.

For positive $\n\leq 2$, $\GG$ cannot grow faster with temperature than the contribution to $H$ from $\rho_R$, cf.~\eqref{TandHdefs}.
 If $T_\n > T_{{\rm re} 0}$, then the temperature-dependent correction to $\GG$ remains smaller than $H$ at all higher temperatures $T>T_{{\rm re} 0}$. Hence, for $\n\leq 2$ the thermal
corrections to $\GG$ can only affect $\Nreh$ if $T_\n < T_{{\rm re} 0}$, which translates into the condition $\Gamma_\n/\Gamma_0 < (\pi^2 g_* /90)^{\n/4}(\frac{m_\phi}{\Gamma_0}\frac{m_\phi}{M_{pl}})^{\n/2}$ to avoid a modification of $\Nreh$.
If $\Gamma_0$ and $\Gamma_\n$ originate from the same interaction, then they tend to have a similar dependence $\Gamma_0, \Gamma_\n \sim \g^2 m_\phi$
on $\g$ and $m_\phi$, cf.~table \ref{ThermalRates}, and this requirement translates into 
$\g\ll\epsilonvarM^{1/2} g_*^{1/4}$,
which is much weaker than the bound \eqref{GeneralScaling} from the absence of a parametric resonance.
We can therefore ignore the impact of thermal feedback on $\Nreh$ for $\n\leq2$, consistent with what was found in \cite{Drewes:2015coa}.
The cases $\n=1$ and $\n=2$ cover several important interactions,  cf.~table \ref{ThermalRates}:
The case $\n=1$ e.g.~describes a Bose enhanced  rate for decays $1\to2$ or $2\to2$ annihilations,
the case $\n=2$ can e.g.~describe scatterings 
amongst scalars.\footnote{For the specific examples given in table \ref{ThermalRates} this conclusion could also have reached without the expansion \eqref{nexpansion} 
by noticing that $f_B(\M_\phi)$ becomes of order unity when $T\sim m_\phi$.}

For $\n>2$ the above argument does not hold. We shall instead use the criterion that $\Gamma_\n (T_{\rm max}/m_\phi)^\n < H$, where $T_{\rm max}$ is the maximal temperature during the reheating epoch.
Using the explicit solutions for \eqref{AveragedEoM} 
found in \cite{Drewes:2014pfa,Ming:2021eut} for 
\begin{eqnarray}\label{initialcondreh}
R = 0 \ , \ 
\Upphi = \rhoend/m_\phi^4 \ {\rm at} \ x=1
\end{eqnarray}
and assuming $T_{{\rm max} 0} > T_\n$ (otherwise thermal effects are certainly negligible), we can approximate
\begin{eqnarray}
T_{\rm max} \simeq m_\phi 
\left(
\frac{30}{\pi^2 g_*}
\right)^{1/4}
\left(
\frac{A_\n}{4}
\right)^{1/(4-\n)}
\left(
\frac{8-2\n}{3}
\right)^{\frac{3}{(\n-4)(5-2\n)}}.
\end{eqnarray}
with 
\begin{eqnarray}
A_\n = A_0 \frac{\Gamma_\n}{\Gamma_0}
\left(
\frac{30}{\pi^2 g_*}
\right)^{\n/4} 
\ , \ 
A_0=\sqrt{2}\frac{\Gamma_0}{m_\phi}\frac{M_{pl}}{m_\phi}\frac{\varphi_{\rm end}}{m_\phi}
\end{eqnarray}
This maximal temperature is reached at 
$x_{\rm max} \simeq \left(\frac{8 - 2n}{3}\right)^{\frac{2}{5 - 2\n}}$. 
We can assume that $\Upphi$ is roughly constant until the moment when $\GG=H$ and hence set $\Upphi\simeq \rho_{\rm end}/m_\phi^4\simeq \frac{2}{3}(\varphi_{\rm end}/m_\phi)^2$. This gives
\begin{eqnarray}
H \simeq \frac{\sqrt{2}}{3} m_\phi \frac{\varphi_{\rm end}}{M_{pl}} x^{-3/2}.
\end{eqnarray}
Using the above expressions 
we can translate the condition that $\GG < H $ at $x=x_{\rm max}$ into 
\begin{eqnarray}
\left(
\frac{\Gamma_\n}{m_\phi}
\right)^{\frac{4}{4-\n}}
< 
\frac{2^{\frac{2+\n}{4-\n}}}{3}
\left(
\frac{8-2\n}{3}
\right)^{
\frac{6(\n-2)}{(2\n-5)(\n-4)}
}
\left(
\frac{30}{\pi^2 g_*}
\right)^{
\frac{\n}{\n-4}
}
\left(
\frac{m_\phi}{M_{pl}}
\right)^{\frac{2\n}{4-\n}}
\left(
\frac{\varphi_{\rm end}}{M_{pl}}
\right)^{\frac{2\n-4}{\n-4}}\label{strongcond}
\end{eqnarray}

This imposes an upper bound on $\Gamma_\n$ from the requirement that thermal effects do not modify the reheating temperature.\footnote{Surprisingly it does not depend on $\Gamma_0$. However, note  that all of this is only relevant if $T_{{\rm max} 0} > T_\n$. This bring ins an additional conditions that depends on $\Gamma_0$.} 
If we assume $\Gamma_\n \propto \g^2 m_\phi$, it turns out that \eqref{strongcond} is weaker than the conditions I) and II) for $\n=1,2,3$.

We have so far assumed that the inflaton is the only field with a non-zero expectation value during reheating. However, in many particle physics models that can accommodate single field inflation the inflaton is not the only scalar field, but simply represents the relevant direction in field space. In fact, leaving aside the specific case of Higgs inflation \cite{Bezrukov:2007ep}, we know for sure that there are at least four real scalar degrees of freedom, the components of the SM Higgs doublet \cite{Ema:2016kpf,Passaglia:2021upk}. 
Even if the dynamics during inflation can effectively be described by a single field model (with $\Phi$ simply being the relevant direction in field space), a certain fraction $\varepsilon$ of the total energy density may be stored in these other degrees of freedom. 
If the time that it takes for this energy to be transferred into $\rho_R$ is much shorter than the value of $\Nreh$ that one would obtain by neglecting this effect, then one can obtain a conservative bound \corr{on} the impact that this has on the actual value of $\Nreh$ by replacing the initial conditions \eqref{initialcondreh}
at $x=1$ with 
$\Upphi = (1-\varepsilon)\rhoend/m_\phi^4$, $R = \varepsilon\rhoend/m_\phi^4$ and using the analytic solutions presented in \cite{Drewes:2014pfa,Co:2020xaf,Ming:2021eut}. If the fraction $\varepsilon$ is subdominant this will not change our conclusions. If either $\varepsilon$ is dominant or its lifetime is longer than that of $\varphi$, a detailed model-dependent study is necessary.

\section{On the consistent description of reheating}\label{ApendixQFT}
A cautious reader may be concerned about the fact that the arguments presented in this work are not derived within a consistent formal description of the reheating process. 
As a result, it may e.g.~not be obvious how the particle production obtained from the classical Mathieu equation \eqref{Mathieu} is related  
to the rates in table~\ref{ThermalRates} that were computed in thermal quantum field theory.
Moreover, the use of an Markovian equation of the form \eqref{EOM} during reheating is in general questionable  because the effective action for $\varphi$ 
after tracing out other degrees of freedom
is fundamentally non-local. 
In this appendix we briefly comment on these points and illustrate how the expressions that we used throughout the main text can be derived consistently from nonequilibrium quantum field theory. For a more complete and pedagogical introduction to the topic we e.g.~refer the interested reader to \cite{Berges:2004yj}.

All properties of a system in quantum statistical mechanics are in principle encoded in the von Neumann density operator $\varrho$. 
Instead of directly dealing with the infinite-dimensional matrix $\varrho$, it is often more convenient to consider a finite set of correlation functions. 
The most important properties of a scalar field $\Phi$
can be encoded in 
its expectation value and the two independent real-valued two-point functions
\begin{eqnarray}\label{SpectralStatistical}
\varphi(x) \equiv\langle\Phi(x)\rangle, 
\
\Delta_\phi^-(x,y)\equiv \ii\langle[\Phi(x),\Phi(y) ]\rangle, 
\
\Delta_\phi^+(x,y)\equiv \frac{1}{2}\langle\{\Phi(x),\Phi(y)\}\rangle - \varphi(x)\varphi(y),
\end{eqnarray}
where $\lbrace \cdot , \cdot \rbrace$ and $[\cdot , \cdot ]$ are the anticommutator and commutator operators respectively.
All other two-point functions of $\Phi$ (such as the Feynman propagator) can be expressed in terms of those defined in \eqref{SpectralStatistical}.
The expectation value is to be taken with respect to both, quantum and thermal fluctuations encoded in the density operator: $\langle\ldots\rangle={\rm Tr}(\varrho \ldots)$.
The one-point function $\varphi$ describes the scalar condensate or \emph{mean field}, while the two-point functions $\Delta^-$ and $\Delta^+$ describe the fluctuations.  
The poles of the \emph{spectral function} $\Delta^-$ encode information about spectrum of (quasi)particles, while the \emph{statistical propagator} $\Delta^+$ contains information about the occupation numbers of each momentum mode. 
For illustrative purposes we restrict the discussion to Minkoswki space, the generalisation to curved space is e.g.~discussed in \cite{Hohenegger:2008zk}.

Our starting point to find the equation of motion for $\varphi$ and $\Delta^\pm$ 
in the closed-time-path (CTP) formalism \cite{Schwinger:1960qe,Bakshi:1962dv,Keldysh:1964ud}
is the two-particle irreducible effective action \cite{Calzetta:1986cq}
with arguments on a suitably defined time path $\mathcal{C}$, 
\begin{eqnarray} \label{2PI-action}
 \mathbf{\GG}[\varphi, \Delta] 
&=& S[\varphi] + \mathbf{\GG}_{\rm loop}[\varphi,  \Delta[\varphi]], \nonumber \\
\mathbf{\GG}_{\rm loop}[\varphi,  \Delta[\varphi]] &=& \frac{i}{2} \text{Tr} \ln \left( \Delta^{-1}\right) +\frac{i}{2} \text{Tr} \left(G^{-1}[\varphi] \,  \Delta\right)
+\mathbf{\GG}_2[\varphi, \Delta],
\end{eqnarray}
with $S[\varphi]$ the classical action and 
$iG^{-1}[\varphi](x_1.x_2)=\frac{\delta^2S[\Phi]}{\delta\Phi(x_1)\delta\Phi(x_2)}|_{\Phi\to\varphi}$.
The trace-terms 
in \eqref{2PI-action} correspond to the one-loop correction to the effective action, which e.g.~gives rise to the Coleman-Weinberg potential \cite{Coleman:1973jx} in the static limit, while $\mathbf{\GG}_2[\varphi, \Delta]$ contains terms with two or more loops. 
The equations of motion for the correlation functions defined in \eqref{SpectralStatistical} are found by functional differentiation,
\begin{eqnarray} \label{EOM-G}
\frac{\delta \mathbf{\GG}[\varphi, \Delta]}{\delta \varphi(x)} = 0 \ , \quad 
\frac{\delta \mathbf{\GG}[\varphi,  \Delta]}{\delta  \Delta(x,y)} =0 \,.
\end{eqnarray}

\subsection{Equations of motion for propagators and particle numbers}

The equations of motion for the propagators $\Delta^\pm$ are obtained from the second equation in~\eqref{EOM-G}. 
In the case of an homogeneous and isotropic system, after a Fourier transformation in the spacial relative coordinate $\mathbf{x}_1-\mathbf{x}_2$, they can be brought into the form
\begin{eqnarray}
		\Big(
		\partial_{t_1}^2  +\W_\phi^2(t_1;\mathbf{k})
		\Big)\Delta_\phi^-(t_1,t_2;\mathbf{k}) 
		&=&-\int_{t_2}^{t_1}dt'\,\Pi_\phi^-(t_1,t';\mathbf{k})\Delta_\phi^-(t',t_2;\mathbf{k})\,, \label{eq:KBrhot} \\
		\Big(
		\partial_{t_1}^2  +\W_\phi^2(t_1;\mathbf{k})
		\Big)
		\Delta_\phi^+(t_1,t_2;\mathbf{k})
		&=&-\int_{t_i}^{t_1}dt'\,\Pi_\phi^-(t_1,t';\mathbf{k})\Delta_\phi^+(t',t_2;\mathbf{k}) \nonumber\\ 
		&&  +\int_{t_i}^{t_2}dt'\,\Pi_\phi^+(t_1,t';\mathbf{k})\Delta_\phi^-(t',t_2;\mathbf{k})\, ,\label{eq:KBFt}
\end{eqnarray}
with $t_i$ the initial time.
Here $\Pi_\phi^{\rm loc}$, $\Pi_\phi^-$ and $\Pi_\phi^+$ are the local, spectral and statistical self-energies, which can be obtained by decomposing the self-energy on the contour\corr{,} $\mathcal{C}$  $\Pi_\phi(x,y)=2i\delta\mathbf{\GG}_2[\varphi, \Delta]/\delta\Delta$\corr{,}  into 
 \begin{equation}\label{eq:spectral_statistical_selfenergy}
    \Pi_\phi(x,y) =-i\Pi_\phi^{\rm loc}(x,x)\,\delta(x-y)_{\mathcal{C}}+\Pi_\phi^+(x,y)-\frac{i}{2}\Pi_\phi^-(x,y)\, \textrm{sign}_{\mathcal{C}}(x_0-y_0),
\end{equation}
and 
\begin{equation}\label{LocalMass}
\W_\phi^2(t;\mathbf{k})=\mathbf{k}^2 + \mass_\phi^{\rm tree}(t)^2 + \int\frac{d^3\mathbf{q}}{(2\pi)^3}\,\Pi_\phi^{\rm loc}(t,t;\mathbf{k})
\equiv \mathbf{k}^2 + \mass_\phi(t)^2
\end{equation}
with $\mass_\phi^{\rm tree}$
the effective tree level mass defined by 
\begin{equation}
iG_\phi^{-1}[\varphi](x_1,x_2)=-(\Box_{x_1}+\mass_\phi^{\rm tree}(x)^2)\delta_{\mathcal{C}}(x_1-x_2).
\end{equation}
The tree level mass $\mass_\phi^{\rm tree}$ differs from $m_\phi$ due to the contribution from the coupling of $\phi$ to the time dependent background $\varphi$.
Together with the local contribution $\Pi_\phi^{\rm loc}$ to the self-energy from forward-scattering 
$\mass_\phi^{\rm tree}$ determines the effective mass $\mass_\phi$ \corr{that appears in} 
the approximation 
\corr{$\disprel_\phi\simeq\W_\phi$} 
in \eqref{LocalMass}
to the full dispersion relation $\disprel_\phi^2$ of screened $\phi$-particles. $\W_\phi^2$ describes a change in the quasiparticle dispersion relation that depends on $\varphi$ and the occupation numbers in the medium, but not on momentum, and therefore can be parameterised by a time-dependent effective mass $\mass_\phi$.
The full dispersion relation
$\disprel_\phi^2$
can be obtained by including $\Pi_\phi^-$
and can be quite complicated (cf.~e.g.~\cite{Aarts:2001qa}); 
it is determined by the poles of the solutions $\Delta^-$ to \eqref{eq:KBrhot} in Wigner space, hence the name spectral function for $\Delta^-$.
The interpretation of the statistical propagator $\Delta^+$
can be illustrated by noticing that there is a contribution $\rho_\phi$ to the total energy density that can be interpreted as the energy in fluctuations $\phi$ (i.e.~particles) and to leading order reads\footnote{
Complete expressions for the energy-momentum tensor derived from \eqref{2PI-action} by varying with respect to the metric can e.g.~be found in \cite{Herranen:2013raa}.
}
\begin{eqnarray}\label{EnergyDensityDeltaPlus}
\rho_\phi \simeq \frac{1}{2} \left(
\partial_{t_t} \partial_{t_2} + \W_\phi^2(t_1)
\right)\Delta^+(t_1,t_2)|)_{t_1=t_2}.
\end{eqnarray}
For the derivation of conditions I) and II) in section \ref{SingleModeOccupation} 
it is  sufficient to consider the leading approximation $\W_\phi$ in \eqref{adiabaticity} and its equivalents for other particle species, i.e., neglecting the RHS of \eqref{eq:KBrhot}. This can be justified by an expansion in the couplings $\{\vv_i,\g_i\}$ and in $\hbar$. 
By making the same approximation in \eqref{eq:KBFt} we obtain an equation of the form
	\begin{equation}\label{KBEKleinGordon}
        (\partial_{t_1}^2 + \textbf{k}^2 +\mass_\phi^2)
		\Delta_\phi^+
		= 0, 
		\end{equation}
		i.e., a Klein-Gordon equation with a time dependent mass $\mass_\phi$. 
This equation for $\Delta_\phi^+$ has the same form as the equation \eqref{ModeEqEta} for the mode functions
and can be brought into the form \eqref{Mathieu} by making the replacements such as \eqref{Akqquadratic} and \eqref{Akqquartic}.
Hence, at leading order in $\hbar$ the solutions of the Mathieu equation \eqref{Mathieu}
describe the behaviour of the statistical propagator $\Delta_\phi^+$, and, through \eqref{EnergyDensityDeltaPlus}, the energy density stored in $\phi$-particles.
The same holds for the $\chi$-particles, cf.~\eqref{ModeEq} with  \eqref{Akforchi} and \eqref{AkHiggsPortal}. 
This in particular justifies using 
the Mathieu equation as a tool to study the non-perturbative  particle production (iv) because \eqref{Mathieu} represents the leading term in the $\hbar$-expansion of the full quantum equation \eqref{eq:KBFt}.

Elementary processes are not included at this level; they are encoded in the self-energies on the RHS of \eqref{eq:KBFt} and contribute at the next order in $\hbar$, cf.~e.g.~\cite{Berges:2002cz}. 
More precisely, one can define 
\begin{eqnarray}\label{GammaDef}
\Gamma_\phi = -\frac{{\rm Im}\tilde{\Pi}_\phi^-(t,\W_\phi)}{\W_\phi}, \quad
\tilde{\Pi}_\phi^-(t,\W_\phi) = \int_{t_i}^\infty dz \Pi^-(t,t-z).  
\end{eqnarray}
The physical interpretation of $\Gamma_\phi$ and $\Delta_\phi^+$ becomes evident in the adiabatic regime, where 
one can approximate \cite{Drewes:2012qw}
\begin{eqnarray}\label{WKB}
\Delta_\phi^+(t_1,t_2;\mathbf{k}) \simeq \frac{\cos\left(
\int_{t_2}^{t_1}dt' \W_\phi(t')
\right)
\exp\left(-\frac{1}{2}\left| \int_{t_2}^{t_1}dt' \Gamma_\phi(t')\right|
\right)
}{2\sqrt{\W_\phi(t_1) \W_\phi(t_2)}}(1 + 2f_\phi(t))
\end{eqnarray}
with $t={\rm min}(t_1,t_2)$. 
Hence, $\Gamma_\phi$ describes the damping of the statistical propagator in time~\cite{Garbrecht:2011xw,Drewes:2012qw}; it determines the rate at which the quasiparticle phase space distribution $f_\phi$ and the energy density \eqref{EnergyDensityDeltaPlus} approach equilibrium.
In the weak coupling and dilute gas limit $f_\phi$ fulfils a standard Boltzmann-equation in which the collision term is computed from \eqref{GammaDef}. 
If the fields that $\Phi$ interacts with are in thermal equilibrium \cite{Anisimov:2008dz},
\eqref{GammaDef} reduces to the well-known statement that the damping is given by the imaginary part (more precisely: discontinuity) of the retarded self-energy $\Pi^R_\phi$ \cite{Weldon:1983jn}, i.e.,
$\Gamma_\phi \simeq   {\rm Im}\Pi^-(\W_\phi)/(2i \W_\phi) = -{\rm Im}\Pi^R(\W_\phi)/\W_\phi$.
Of course, the same considerations hold for the spectral and statistical propagators for all other fields $\X$.

\subsection{Equation of motion for $\varphi$}\label{EOMvarphi}

The equation of motion for $\varphi$ can be obtained from \eqref{2PI-action} 
via the first equation in~\eqref{EOM-G}. 
Assuming that we have formally solved the equations of motion
\eqref{eq:KBrhot} and \eqref{eq:KBFt}
for the propagators in the background provided by $\varphi$, we can use the solutions $\Delta[\varphi]$ to compute the loops in $\mathbf{\GG}_{\rm loop}[\varphi,  \Delta[\varphi]]$.
Since we are interested in oscillations near the minimum, we can perform a Taylor expansion in functional space around the ground state configuration $\bar{\varphi}$
as $\varphi = \bar{\varphi} + \delta\bar{\varphi}$. 
For simplicity we shall assume that the ground state corresponds to the field configuration $\bar{\varphi}=0$ and write \cite{Buldgen:2019dus}
\begin{align}\label{FunctionalTaylor}
   \left.\frac{\partial \mathbf{\GG}_{\rm loop}[\varphi,  \Delta[\varphi]]}{\partial \varphi(x)}\right|_{
\substack{\varphio +  \delta\bar{\varphi}} }
& = \left. \frac{\partial \mathbf{\GG}_{\rm loop}[\varphi,  \Delta[\varphi]]}{\partial \varphi(x)} \right|_{\varphio}\nonumber \\
  + \sum_{n=1}^\infty \frac{1}{n!}
 \prod_{i=1}^n & \left[\int_{\mathcal{C}}  dx_i^0  
 \int d^3 \x_i \; \delta\bar{\varphi}(x_i)\right]\left[
 \frac{\delta^n}{\delta \varphi(x_1) \cdots \delta\varphi(x_n)}\left(\left.\frac{\partial\mathbf{\GG}_{\rm loop}[\varphi,  \Delta[\varphi]]}{\partial \varphi(x)}
 \right)\right]\right|_{\bar{\varphi}}.
\end{align}
where the partial functional derivatives $\partial / \partial \varphi$ only apply to the explicit dependence of $\mathbf{\GG}_{\rm loop}$. 
A Markovian (local) equation  can be derived 
in the regime $\dot{\varphi}/\varphi < \Gamma_\X$ \cite{Berera:2008ar,Mukaida:2013xxa,Cheung:2015iqa,Buldgen:2019dus}
and in the mildly non-linear regime \eqref{MildlyNonlinear} 
\cite{Calzetta:1989vs,Greiner:1996dx,Mukaida:2013xxa,Ai:2021gtg}.
Being interested in reheating, we focus on the latter case. 
In this mildly nonlinear regime \eqref{MildlyNonlinear} we may neglect the $\varphi$-dependence in the propagators $\Delta[\varphi]$ inside the loops, which would typically break the adiabaticity condition \eqref{adiabaticity} outside the regime \eqref{MildlyNonlinear}.
We can further locally 
neglect the effects of  nonlinearities  
inside memory kernels in \eqref{FunctionalTaylor} and approximate \cite{Greiner:1996dx}
\begin{equation}
\varphi(t') = \varphi\bigl( t \!-\! (t\!-\!t') \bigr)
	\approx
	\varphi(t) \cos\bigl[ \M_\phi(t\!-\!t') \bigr]
	- \frac{\dot{\varphi}(t)}{\M_\phi} \sin\bigl[ \M_\phi(t\!-\!t') \bigr].
\label{VarphiExpansion}
\end{equation}
One then obtains an equation of the form
\begin{equation}
\mathcal{K}(\varphi) \ddot{\varphi} 
	+ \frac{1}{2} \mathcal{K}'(\varphi) \dot{\varphi}^2
	+ \GG(\varphi) \dot{\varphi}
	+ \mathcal{V}'(\varphi) = 0 \, .
\label{proper local eq}
\end{equation}
We note that there is no noise-term on the RHS of \eqref{proper local eq} and \eqref{EOM}, as one might expect from a generalised fluctuation-dissipation theorem, because our definition of $\varphi$ includes an average over thermal fluctuations. Near the potential minimum, where reheating occurs, these average to zero. The equation of motion for $\Phi$ itself would contain a noise term, see e.g. \cite{Chou:1984es,Morikawa:1986rp,Greiner:1998vd,Yokoyama:2004pf,Boyanovsky:2004dj,Anisimov:2008dz,Cheung:2015iqa}.

In \cite{Ai:2021gtg} it has been shown explicitly that the use of Markovian equations of motion of the form \eqref{proper local eq} based on the approximation \eqref{VarphiExpansion} is justified in the mildly non-linear regime, were deviations from \eqref{MildlyNonlinear} can be treated as perturbations.
The coefficient $\mathcal{K}(\varphi)$ does not lead to dissipative behaviour, but causes a perturbative correction to the effective frequency $\M_\phi$.
While the impact of this on the phase of the oscillations can build up after many oscillations, it modifies $\GG$ only 
by shifting the mass shell of the quasiparticles that compose the condensate $\varphi$ away from $m_\phi$ to $\M_\phi$. 
While this kinematic effect in principle can be significant \cite{Kolb:2003ke,Yokoyama:2005dv,Bodeker:2006ij,Drewes:2010pf,Mukaida:2012qn,Mukaida:2012bz,Drewes:2013iaa,Drewes:2014pfa,Drewes:2015eoa}, it is always a small correction in the regime allowed by conditions I) and II), cf.~section \ref{ThermalBrief}.\footnote{
More precisely: In the mildly non-linear regime, where the field elongation is restricted by \eqref{MildlyNonlinear} and occupation numbers in the primordial plasma are restricted by the considerations in section \ref{ThermalBrief}, corrections of higher order in the loop expansion are suppressed by additional powers of the coupling constants, which are restricted to small values by the conditions \eqref{GeneralScaling}. This indicates that higher order corrections are unobservably small during reheating. }
Since we are not interested in the phase, but only in the moment when $\GG=H$, we can neglect the correction $\mathcal{K}(\varphi)$ and
conclude that the use of the equation \eqref{EOM} is consistent in the mildly non-linear regime of elongations permitted by the considerations in section \ref{sec:selfinteractions}.

This still leaves the question how $\GG$ is computed practically. The operators considered in sections \ref{ALPSec} and \ref{YukawaSec} as well as the case $n=1$ in section \ref{OtherScalarSec} are linear in $\Phi$. 
In this case $\GG$ is (at leading order) independent of $\varphi$, meaning that the only non-linearities in \eqref{proper local eq} are due to deviations from \eqref{MildlyNonlinear} in $\V(\varphi)$, which we know must be small. In this case, 
it is well-known that the leading contribution to $\GG$ can be identified with the thermal width $\Gamma_\phi$ of the propagator in \eqref{GammaDef} because the frequency at which $\varphi$ oscillates coincides with the pole mass of $\phi$-particles, as illustrated before \eqref{varphicdef}.
Higher order corrections are suppressed by additional small parameters, cf.~footnote \ref{FootnoteHarmonicExpansion}.
In this situation the inflaton damping rate $\GG$ and the rate of particle production are both identical to $\Gamma_\phi$ in \eqref{GammaDef} and given by table \ref{ThermalRates}.\footnote{
A careful reader may be concerned about the fact that 
the conditions I) and II) in section \ref{SingleModeOccupation} are formulated in terms of the tree-level Mathieu equation \eqref{Mathieu}, neglecting contributions to $\Gamma_\varphi$ from elementary decays that appear as radiative corrections to the equations of motion in the CTP formalism,
while one-loop contributions are later used to compute the coefficients in table \ref{ThermalRates}. 
This is, however, not inconsistent. 
Using \eqref{Mathieu} based on the approximation \eqref{KBEKleinGordon} leads to a conservative estimate of the parameter range where 
feedback due to non-perturbative effects at tree-level can be avoided
(because elementary decays encoded in one-loop corrections on the RHS of \eqref{eq:KBFt} 
can weaken and possibly avoid resonant particle production by reducing the occupation numbers between two zero-crossings of $\varphi$). 
Within that parameter range one can then be confident that non-perturbative effects at tree-level are sub-dominant, and therefore the leading contribution to $\Gamma_\varphi$ comes from elementary processes encoded in one-loop diagrams.
  Hence, it is fully consistent to neglect the loop corrections when formulating conditions I) and II), and then neglect the tree-level effects when computing $\Gamma_\varphi$ within the range of parameters where these conditions are fulfilled.
} 
When $\GG$ depends on $\varphi$ (as e.g.~for $n>1$ in section \ref{OtherScalarSec}) the connection between $\GG$ and the rate of perturbative particle production is more subtle (cf.~section 5 in \cite{Ai:2021gtg}), but the expression in table \ref{ThermalRates} can still be used. 
\end{appendix}
\bibliographystyle{JHEP}
\bibliography{InflatonCMB}{}

\end{document}